\documentclass{statsoc}
\usepackage[a4paper]{geometry}

\usepackage[utf8]{inputenc}
\usepackage{lipsum}
\usepackage{xcolor}
\usepackage{etoolbox}

\makeatletter
\patchcmd{\@makecaption}
  {\parbox}
  {\advance\@tempdima-\fontdimen2} % decrease the width!
  {}{}
\makeatother  

\usepackage[caption = false]{subfig} 
\usepackage{enumerate}
\usepackage{comment}
\usepackage{natbib, float}

\usepackage{graphicx}
\usepackage{tikz}
\usepackage{pgf}
\usepackage{pgflibraryarrows}
\usepackage{amsmath}

\usepackage{amsfonts}
\usepackage{amssymb}
\usepackage{url}
\DeclareSymbolFont{bbold}{U}{bbold}{m}{n}
\DeclareSymbolFontAlphabet{\mathbbold}{bbold}

\title[Bayes ALAAM]
{Bayesian Analysis of Social Influence}

\author[Koskinen and Daraganova]{Johan Koskinen}
\address{Melbourne School of Psychological Sciences, Faculty of Medicine, Dentistry, and Health Sciences, University of Melbourne, Australia\\ Institute of Analytical Sociology, University of Link{\"o}ping, Sweden
}
\email{jkoskinen@unimelb.edu.au}
\author[Koskinen and Daraganova]{Galina Daraganova}
\address{Business Intelligence, South Eastern Melbourne Primary Health Network}

\begin{document}
\maketitle
\begin{abstract}
The network influence model is a model for binary outcome variables that accounts for dependencies between outcomes for units that are relationally tied. The basic influence model was previously extended to afford a suite of new dependence assumptions and because of its relation to traditional Markov random field models it is often referred to as the auto logistic actor-attribute model (ALAAM). We extend on current approaches for fitting ALAAMs by presenting a comprehensive Bayesian inference scheme that supports testing of dependencies across subsets of data and the presence of missing data. We illustrate different aspects of the procedures through three empirical examples: masculinity attitudes in an all-male Australian school class, educational progression in Swedish schools, and unemployment among adults in a community sample in Australia.
%We apply the model to three datasets to illustrate different aspects of the procedures. For male students in an Australian school we find that attitudes to masculinity demonstrate clear peer-dependencies, where having friends with higher than average masculinity attitudes is associated with having a higher masculinity attitude yourself. For a students in Swedish schools, there is evidence for network effects in the intention to proceed to higher education. For a sample of a community network in Australia, we find that unemployment tends to cluster among people that are connected. 
\end{abstract}
\keywords{Auto-logistic actor attribute model; Exponential family models; Ising model; Contagion; Social Influence; peer effects}

\section{Introduction}
%{\bf{generalised exchange Schaefer?}}
In social statistics it has become commonplace to take dependencies between outcomes into account using multilevel models (e.g. Goldstein, 1995; Snijders and Bosker, 2011). Thus, when we consider educational outcomes, we may account for compositional or contextual factors with random effects for school classes or neighbourhoods. If we acknowledge the possibility of our observational units being connected with each other through social networks, we can account for some of the dependence this induces using multilevel models (Tranmer et al., 2014) but we cannot capture the detail and diversity of what the social networks literature has termed \emph{social influence} (Robins, 2015). The notion of attitudes and information spreading through friendship networks was already a premise in Moreno's (1934) seminal work explaining a runaway epidemic in a reformatory. Coleman, Katz, and Menzel's (1957) study of the diffusion of the prescription of a novel drug among a network of physicians has been followed by numerous empirical studies of spread on different types of networks (for example Strang, 1991; 
%Strang and Soule, 1998;
Strang and Tuma, 1993; Valente, 1995
%, 2005, 2012
). Here, one of our examples aims to investigate if a young, male student's masculinity attitude, coded as high (1) or low (0), depends on whether their friends' attitudes are high or low.

Social influence in social network analysis can broadly be seen as representing processes whereby people tend to be, or become, similar to their friends (or contacts) in their behaviours, attitudes, or beliefs. Social influence is sometimes referred to as \emph{social contagion} (e.g. Robins et al., 2012, Burt, 1987) by analogy to how diseases spread through contact between individuals. Network models are indeed frequently used to model disease spread and epidemics (Morris, 2004; Rolls et al. 2012%, 2013a, 2013b
; Jenness, Goodreau, Morris, 2016; Krivitsky and Morris 2017) even though the mechanisms of social contagion may differ. The current canonical empirical framework for investigating social influence is stochastic actor-oriented models, SAOMs (Steglich et al., 2010). While a powerful tool, SAOM require longitudinal network data and researchers do not always have the resources or opportunity to collect network data at multiple points in time. For cross-sectional network data, even if you have to assert the existence of contagion or influence, controlling for dependencies is a statistical reality (see e.g. Bailey and Hoff, 2015) and neglecting these dependencies may have adverse effects (Doreian, Teuter, \& Wang, 1984; Lubbers \& Snijders, 2007). Consequently, we defer to other work for discussions and analysis of identification of peer effects (Manski, 1993; An, 2011; Bramoull\'{e}, Djebbari, and Fortin, 2009) and focus here on the inferential aspects of a well-defined framework for accounting for network dependence in individual outcomes.

We consider a class of models for investigating social influence for cross-sectional data called auto-logistic actor-attribute models (ALAAMs) (Robins et al., 2001; Daraganova and Robins, 2013) where the outcome of interest is binary. A number of continuous models for social influence exist (Marsden \& Friedkin, 1994; Leenders, 2002; Doreian, 1982; Agneessens and Koskinen, 2016; Sewell, 2017; Vitale et al., 2016) that can easily be modified to suit binary outcome variables (Koskinen and Stenberg, 2012; Zhang et al., 2013) but these do not allow specifying the types of dependencies that the ALAAM does.
% These dependencies stem from how ALAAM relate to Gibbs random fields. 

%A review of causal effects for peer effects in a network context (); 

Gibbs random fields, such as the auto-logistic Ising model (Besag, 1972), have been studied in great detail in statistics and employed in various forms in spatial statistics for modelling binary outcomes with neighbourhood dependencies. To accommodate interpretations in terms of the Behavioural and Social Sciences, Robins et al. (2001) elaborated on these Gibbs-distributions and derived a class of `social influence' models from a set of specific dependence assumptions.
These were later extended by Daraganova (2009) to form a family of actor-attribute auto-logistic models for inferring contagion in cross-sectional data. Exponential random graph models (ERGMs) is a related class of models aimed at modelling the network ties conditional on fixed actor covariates (see Lusher et al., 2013, for an introduction). There are known problems with ERGMs (Handcock, 2003; Schweinberger, 2011) and it is well-known that
simple model-specifications do not work, for example in the sense that the maximum likelihood does not exist or predictive distributions place most of their mass on empty or complete graphs (Snijders et al., 2006; Schweinberger, 2011; see also Section 3.1 of Schweinberger et al., 2020).
For ALAAMs this is less of an issue but inhomogeneous ALAAMs still present considerable challenges relative to the simpler Ising model.

Maximum likelihood estimation for the elaborated model as described in Daraganova and Robins (2013) is implemented in the statistical software package MPnet (Wang et al., 2014) and is becoming increasingly more popular (some recent studies include acquisition of norms through networks, Kashima et al., 2013; network effects on performance, Letina, 2016; and `contagion' of depression and PTSD, Bryant et al., 2017; see Parker, Pallotti, and Lomi, 2021, for a review).
Maximum likelihood estimation follows Snijders' (2002) implementation of the Robbins-Monro (1951) algorithm and the square roots of the diagonal elements of the inverse Fisher information matrix are used as standard errors. These standard errors are motivated by the usual (large $n$) asymptotics for exponential families, asymptotics that do not apply for ERGMs (Schweinberger et al., 2020). There is no reason to assume that the required asymptotics apply for ALAAMs either which means that a Bayesian inference procedure offers considerable advantages over the maximum likelihood approach, the latter not having well defined measures of uncertainty. Similarly, whereas the asymptotics required for, for example, Akaike's Information are not available (nor are the number of observations defined), Bayesian model selection criteria are well-defined. In addition, the Bayesian approach offers a flexible framework for handling missing data and lends itself to extensions to hierarchical modelling.

M{\o}ller et al. (2006) proposed an auxiliary variable MCMC for Bayesian inference for auto logistic models. While this works well for the Ising model, it fails fails for the inhomogeneous ALAAM of the more elaborate model of Robins et al. (2001), and modified MCMC samplers are required (Koskinen, 2008). To accommodate the challenges presented by realistic ALAAM specifications with multiple covariates, we draw on an adoption of the exchange algorithm (Murray et al., 2006) that has previously been applied by Caimo and Friel (2011) to exponential family random graph models. This is an improvement on the previous Bayesian inference approach for ALAAMs (Koskinen, 2008) and provides a straightforward and flexible inference scheme. We demonstrate how this inference procedure caters to the practical issues often encountered when working with complex empirical network data, such as handling missing data, performing goodness-of-fit, and choosing between competing models. We introduce the model by describing it in some detail. We then proceed to outline various aspect of inference for the model, something that we then illustrate in three empirical examples.

\section{Notational preliminaries}
We consider networks represented as graphs or  digraphs $G(V,E)$ on a fixed set of nodes $V= \{ 1,2,\ldots, n \}$, with an arc-set $E \subset V^{(2)}=\{ (i,j) \in V \times V :i \neq j \}$, for digraphs, and $E \subset { V \choose 2}$ for graphs.
In social network research, the nodes typically represent $n$ individuals and $E$ the set of connections amongst them (Robins, 2015). We further assume a stochastic binary vertex labelling $\varphi:V \rightarrow \{0,1\}$, that corresponds to the binary outcome variable of interest for the nodes of the graph. We represent $G ( V,E ,\varphi )$ by its $n \times n$ binary adjacency matrix $X =(X_{ij} : (i,j) \in V^{2} ) $, where the tie-indicators 
\begin{equation*}  
X_{ij}= \left\{
\begin{array}{lr}
	1,&\text{if there is a tie from } i \text{ to } j  \text{ in } G\\
	0,&\text{else}
\end{array} 
\right. {\text{,}}
\end{equation*}
and the attribute indicators $Y=(Y_i)_{i=1}^n$
\begin{equation*}  
Y_{i}= \left\{
\begin{array}{lr}
	1,&\text{if } \varphi(i) = 1\\
	0,&\text{else}
\end{array} 
\right. {\text{.}}
\end{equation*}
We denote the space of all adjacency matrices by $\mathcal{X}=\{0,1\}^{V^{(2)}}$ and the support of the attribute vector by $\mathcal{Y}=\{0,1\}^V$. We allow for binary and continuous, fixed and exogenous covariates, but suppress the notational dependency on these for the sake of exposition.

%{\textbf{plain language example of inferential and modelling problem: assume that we are investigating employees approval of SOMETHING, how determine if people that collaborate are more likely to share the same attitude than people who do not collaborate}}

 In the examples to follow in Section~\ref{sec:applications}, $V$ consists of 108 males in a Year 10 level Australian secondary school; 403 sixth grade students across 19 school classes in Sweden; and 551 adult individuals in Australia. For the first two cases, the network ties $X$ are friendship nominations (both directed) and for the third, nominations of whom you are close to and/or with whom you discuss employment matters (treated as undirected). The outcome variables ($Y$) are a binary masculine attitudes index, progression to higher education (intention), and employment status, respectively.    

\section{The auto-logistic actor attribute model}
The general form of the log-linear model used here is
\begin{equation}
p_{\theta} (y|x) = \Pr(Y=y| X=x,\theta)= \exp\{ \theta^{\top}z(y,x)-\psi(\theta) \}{\text{,}}
\label{eq:alaam}
\end{equation}
where $z(\cdot)$ is a $p \times 1$ vector-valued function on $G$, $\theta \in \mathbb{R}^p$ are the natural parameters, and
\[
\psi(\theta)=\log\sum_{y\in\mathcal{Y}} \exp\{ \theta^{\top}z(y,x)\}{\text{,}}
\]
is a normalising constant. In the next section we proceed to define dependence assumptions from which the statistics $z(\cdot)$ may be derived.

% Section 3.1 dependence
\subsection{Dependence}
The simplest form of an ALAAM is a model in which $Y_{i}$ and $Y_{j}$ are independent conditional on $X$ and a fixed set of exogenous covariates, for all $i,j \in V$. In this case the ALAAM reduces to a logistic regression model. Auto-logistic models relax the assumption of independence by allowing the state of sites to depend on the states of their \emph{neighbours} in, for example, a lattice like in the Ising model (Besag, 1972).
Besag (1974) elaborate auto-logistic models for different types of lattice systems and define dependencies of the first as well as the second order. The neighbourhood for lattice systems is straightforwardly given by the index set of the site variables. In a rectangular lattice, the variable $y_{i,j}$ has neighbours $y_{i-1,j}$,$y_{i+1,j}$,$y_{i,j-1}$, and $y_{i,j+1}$.  As social networks rarely are regular graphs, nodes will differ both in the number of neighbours they have as well as their structural position within a graph. This makes for a possibly rich dependence structure but it is not self-evident that just taking the observed network as representing the neighbourhood structure makes for a coherent probability model (especially in the case of directed graphs that are not chordal and that have cycles). Besag (1974) briefly discuss how to define neighbourhoods for non-lattice systems such as points distributed in the plane but this offers little advice for general structures such as networks.

Frank and Strauss (1986) derived a class of exponential family models for the network ties $X$ from dependence assumptions represented by a dependence graph. Robins et al. (2001) similarly specified a dependence graph for the variables $Y$ conditionally on $X=x$. In the dependence graph, the absence of a tie between two variables means that the two variables are conditionally independent. Throughout, we will aim to model the conditional probability structure of $Y$ given $X=x$ and make no statement about the marginal probability $\Pr(X=x)$. To capture the fact that $X$ are assumed to be exogenous and $Y$ endogenous, we define a two-block \emph{chain graph} (Wermuth and Lauritzen, 1990) $\mathcal{D}$, with a block consisting of parent variables $X$ and a block consisting of child variables $Y$, and where there may be directed ties from nodes in the parent block to nodes in the child block, and possibly undirected ties between variables within the same block. For the purposes of describing the dependence structure we will use a relabelled index set $T=\{n+1,\ldots,n+t\}$ for the tie-variables whenever there is no ambiguity, where $t=n(n-1)$ or $t=n(n-1)/2$ according to whether the network is directed or not respectively. It is convenient to denote the whole set of variables by $M=(M_i)_{i \in S}$, where $M_i=Y_i$ for $i \in V$ and  $M_i=X_i$ for $i \in T$, $S = V \cup T$. Lower case $m$ is taken to be the corresponding realisation of $M$.

There is a directed tie from a parent node $X_{i}$, $i \in T$, to a child node $Y_h$, if  $X_{j}$ is a parent of $Y_h$, denoted $X_{i} \in \mathrm{pa}(Y_h)$. We say that $X_{i} \in \mathrm{pa}(Y_h) $ if the functional form $\Pr(Y_h=y_h|X=x,Y_{-h}=y_{-h})$ depends on $x_{i}$ and define mutual conditional dependence among child variables $Y_i$ and $Y_j$ as occurring if the functional form of $\Pr(Y_i=y_i|Y_{-i}=y_{-i},X=x)$ depends on $y_j$ and if the functional form of $\Pr(Y_j=y_j|Y_{-j}=y_{-j},X=x)$ depends on $y_i$.

The moral graph $\mathcal{Q}$ of a chain graph $\mathcal{D}$ is obtained by adding undirected edges between parents of the same child and tuning all directed ties in $\mathcal{D}$ undirected. Writing  $Q(m)=\log\Pr(Y=y \mid X=X)-\log\Pr(Y=0 \mid X=X)$, following our definition of $\mathcal{D}$ and Besag's (1974) application of the Hammersley-Clifford theorem, $Q(y)=\sum_{A}\gamma_{A \subseteq S} \prod_{i \in A} M_i$, where $\gamma_A$ are non-zero if and only if $A$ is a clique in $\mathcal{Q}$ (see Section A of the Appendix for further details). This is of the form of  Eq.~\ref{eq:alaam} with statistics that have elements $z_A =\prod_{i \in A} Y_i \prod_{jk \in A} X_{jk}$.

We proceed to describe three basic classes of dependence assumptions that imply three basic types of models. These models are characterised by their own specific set of statistics.

\begin{figure}

\begin{center}
\begin{tikzpicture}
\begin{scope}
\boldmath \tikzstyle{every
node}=[x=3ex,y=2.5ex,shape=circle,minimum size=2ex]
\tikzstyle{every path}=[very thick, -stealth', shorten <=2pt,
shorten >=2pt]

\node (a) at (0,10) {Network Block} ; 
\draw[black,rounded corners] (-1, -1) rectangle (1.5, 3.5) {};
\node[draw,thin] (xij) at (0,6) {{\small $X_{ij}$}} ; 
\node[draw,thin]  (xik) at (1,3) {{\small $X_{ik}$}}; 
\node[draw,thin]  (xil) at (0,0) {{\small $X_{i\ell}$}} ;

\node (b) at (6,10) {Attribute Block} ; 
\draw[black,rounded corners] (2, 0) rectangle (4, 3) {};
\node[draw,thin] (yi) at (6,3) {{\small $Y_{i}$}} ; 

\draw (xij) -> (yi);  \draw (xik) -> (yi);  \draw (xil) -> (yi);

\node[draw,thin] (bxij) at (14,6) {{\small $X_{ij}$}} ; 
\node[draw,thin]  (bxik) at (15,3) {{\small $X_{ik}$}}; 
\node[draw,thin]  (bxil) at (14,0) {{\small $X_{i\ell}$}} ;

\node[draw,thin] (byi) at (20,3) {{\small $Y_{i}$}} ; 

\draw[-] (bxij) -- (byi);  \draw[-] (bxik) --(byi);  \draw[-] (bxil) -- (byi);  
\draw[-] (bxij) -- (bxik);  \draw[-] (bxik) -- (bxil);  \draw[ -stealth', shorten <=2pt, shorten >=2pt, -]  (bxij) to [bend right] (bxil); 

\node (c) at (0,-3) {(a)} ; 
\node (c) at (14,-3) {(b)} ;

\end{scope}
\end{tikzpicture}\\
\end{center}
\caption{Dependence graph (a) and Moral graph (b) of network activity dependence model (Robins et al., 2001), where $X_{ij}$ are tie variables and $Y_i$ binary nodal outcomes, for $i,j \in V$}\label{fig:dep1}
\end{figure}
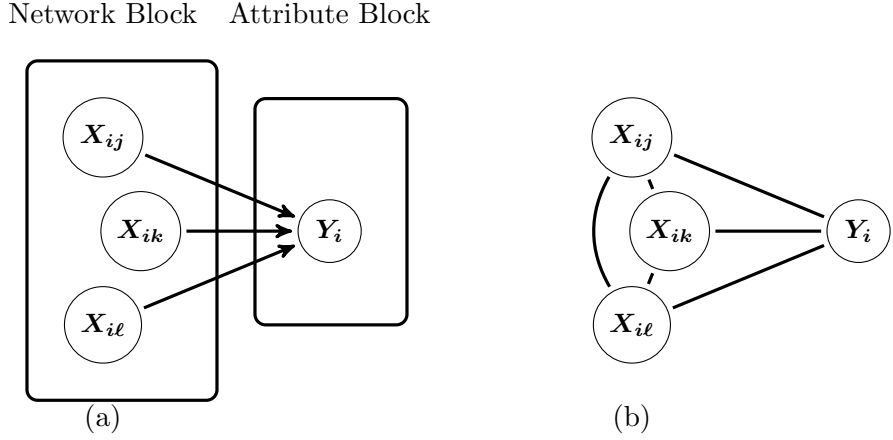

\subsubsection{Network Activity Dependence}
The simplest form of ALAAM that still accounts for some dependence on the network ties leads to a model where outcomes are conditionally independent, conditionally on the network. Robins et al. (2001) followed the method of Frank and Strauss (1986) in defining a dependence graph based on the elements of the index set $V$ of $Y$ and $X$. They define the ``Network Activity'' [NA] dependence assumption.

\emph{Assumption}[NA]: An attribute variable $Y_i$ is conditionally dependent on the network tie-variable $X_{kh}$ if and only if $\{i \}\cap \{k,h \}\neq\emptyset$.

The [NA] dependence assumption defines the chain graph (a) in Figure~\ref{fig:dep1} whose moral graph is (b) in Figure~\ref{fig:dep1}. Cliques in $\mathcal{Q}$ are the singletons of type $Y_i$, two-cliques of the type $(X_{ij} , Y_i)$, $(X_{ik} , Y_i)$, and  $(X_{ih} , Y_i)$, as well as three-cliques corresponding to labelled network 2-stars $(X_{ij} ,  X_{ik} , Y_i)$ and four-cliques corresponding to labelled network 3-stars $(X_{ij} ,  X_{ik}, X_{ih} , Y_i)$, as well as higher order cliques, all corresponding to the association of different stars with the attribute value $y_i=1$.
 The model  (\ref{eq:alaam}) thus has statistics of the form $y_{i}x_{i,k_1}\cdots x_{i,k_s}$ with parameters $ \theta_{i,k_1,\ldots,k_s}$, for $i \in V$ and $s$-element subsets $(k_1,\ldots,k_s) \in {V_{-i} \choose s}  $. To reduce the number of parameters, the following homogeneity assumption may be imposed:
\[
\theta_{i,k_1,\ldots,k_s}=\theta_{s}
\]
for all $i \in V$ and $(k_1,\ldots,k_s)\in {V_{-i} \choose s} $. Thus, for example the interaction terms $X_{ik}X_{ij}Y_{i}$ and $X_{jk}X_{ji}Y_{j}$ both contribute to the same sufficient statistic (see Koskinen et al., 2018, on homogeneity constraints in ERGMs).

While [NA] takes the dependence on network ties into account, the nodal outcomes are conditionally independent for all $i \in V$ conditional on $X=x$. Thus, while the dependence assumption of Frank and Strauss (1986) for the conditional model of $X$ given $Y$ induces dependencies, [NA] does not induce dependence for the elements of $Y$ given $X$. In the network activity model, the statistic $\sum_i Y_i$ acts as regular intercept term. The \emph{activity} statistic,  $\sum_i Y_i X_{i+}$ informs us of the extent to which nodes that have a high degree (are popular) are more or less likely to have a non-zero outcome on $Y$. The two-stars statistic $ \sum_i Y_i {X_{i+} \choose 2}$ effectively acts as a quadratic degree effect. Higher order star statistics may similarly be interpreted as various forms of accounting for the effect on the outcome of heterogeneity in the degree distribution. For directed networks, the network activity dependence model allows for a richer description of the dependence of the outcome on network structure. In addition to the baseline effect of degree, the model includes a reciprocity statistic $\sum_{i}Y_i\sum_{j \neq i}X_{ij}X_{ji}$ which may capture the extent to which nodes that have many reciprocated ties are more or less likely to have the outcome. Among directed stars, we may have out-2-stars $Y_iX_{ij}X_{ik}$; in-2-stars $Y_iX_{ji}X_{ki}$; as well two-paths  $Y_iX_{ij}X_{ki}$. The two-stars have relevance from an influence perspective to the extent that they relate to nodes that are in-between other nodes, acting as brokers (Burt, 1987).

\subsubsection{Network Contagion Dependence}
Modelling social influence requires us to define a model that relaxes independence of outcomes for actors.  A number of different network social influence mechanisms have been specified (Friedkin, 1984; Marsden and Friedkin, 1994; Burt, 1987). Here we define network contagion in terms of dependence between $Y_{i}$ and $Y_{j}$ for pairs $(i,j)\in E$. We cannot simply represent this in the chain graph by adding edges between child variables. For modelling network ties, Pattison and Robins (2002) defined partial conditional dependence in order to extend the local dependence assumptions of Frank and Strauss (1986) by allowing the dependence between variables to be contingent on the states of a third variable. Daraganova (2009) elaborated a number of partial conditional dependence assumptions for the ALAAM that define a number of contagion parameters. We call the first dependence assumption ``Direct Contagion'' denoted as [DC].

\emph{Assumption}[DC]: Any two attribute variables $Y_i$ and $Y_j$ are conditionally dependent if and only if they are connected by a tie $x_{ij}=1$.

This dependence assumption can be represented by a series of partial dependence structures $\mathcal{Q}_{B}$ for subsets of variables $B  \subseteq S$. The node set of  $\mathcal{Q}_{B}$  is $S \backslash B$ and $\{i,j\} \in \mathcal{Q}_{B}$ if $M_i$ and $M_j$ are conditionally dependent, given that $M_h=m_h$ for $h\in S \backslash \{i,j\}$ and $M_h=0$ for $h \in B$. Consequently, for [DC], $\{Y_i.Y_j\}$ is not an edge of  $\mathcal{Q}_{B}$ for $B=S \backslash \{ Y_i, Y_j \}$, but  $\{Y_i,Y_j\}$ may be an edge of  $\mathcal{Q}_{B}$ for $B=S \backslash \{ Y_i, Y_j , X_{ij}\}$. The partial dependence structures prescribe what interactions among variables $S$ are zero in (\ref{eq:alaam})(see Section B of the Appendix for further details): 

If $A \subseteq S \backslash B$ and $A$ is not a clique in $\mathcal{Q}_B$ for some $B \subset S$, then the parameter $\theta_A$ corresponding to the statistic $\prod_{i \in A}M_i$ is $0$. It follows that the parameter  $\theta_A$  is non-zero if and only if $A$ is a clique in $\mathcal{Q}$  and in all $\mathcal{Q}_B$ for which $A \cap B = \emptyset$. 

Thus, every possible outcome $x \in \mathcal{X}$ determines a dependence structure described by the moral graph $\mathcal{Q}$ and the associated sequence of partial dependence structures $\mathcal{Q}_B$. Non-zero interactions in (\ref{eq:alaam}) under assumption [DC] include the non-zero interactions of the network activity model as well as the contagion statistic $\sum_{i < j}Y_iY_jX_{ij}$. A positive parameter for the contagion statistic means that a node is more likely to have the outcome if it is connected to another node that has the outcome. Note that the cliques corresponding to non-zero interactions are not necessarily hierarchical. For example, while $Y_iY_jX_{ij}$ may be a non-zero interaction,  $Y_iY_j$ is not for $\{i,j\} \subset B$. Further, a model with the activity star statistics and the contagion term satisfies [DC] but so does a model with additional interactions. The choice of what parameters to set to zero thus involves decisions that, albeit arbitrary, can be motivated from the perspective of parsimony. For the direct contagion model, it would be wise to set all interactions involving more than two outcome variables to zero. Directed ties in the network allow for more elaborate network effects and network contagion effects but care has to be taken to respect the symmetry of the mutual dependencies among child variables (see Section E of the Appendix for further details).

\subsubsection{Indirect Network and Contagion Dependencies}
Dependencies of outcomes can be further elaborated to incorporate nodes at distances greater than one.
We can allow for dependencies of outcome variables on indirect ties. A first assumption, ``Indirect Structural Influence'' [ISI], elaborates on the structural effects of [NA].

\emph{Assumption} [ISI]: An attribute variable $Y_i$ is conditionally dependent on any network tie-variable $X_{hk}$. $h,k \neq i$, if and only if
 $x_{ik}=1$ or $x_{ih}=1$.

The assumption [ISI] affords statistics that capture how being connected to nodes that themselves have many ties (are popular) may affect the outcome as well as the effect of triadic closure (see Figure 13(a) in the Appendix). We may also allow for dependence of the outcome of $Y_i$, on outcomes $Y_j$, of those nodes $j$ to whom node $i$ is only indirectly connected, in order to capture a form of indirect contagion (Brock \& Durlauf, 2002; Marsden \& Friedkin, 1993) through the ``Indirect Dependent Attribute'' [IDA] assumption.

\emph{Assumption} [IDA]: Any two attribute variables $Y_i$ and $Y_k$ ($i \neq k$) are conditionally dependent if and only if they are directly connected or connected by a path of length two,   $x_{ij}=x_{jk}=1$, for some $j \neq i,k$. 

In a regular lattice, [IDA] yields the second-order neighbourhoods of Besag's (1974) auto-logistic model for Plantago Ianceolata but [DC], [ISI], and [IDA] taken together and applied to general network structures yield an extensive set of possible non-zero interactions. Non-zero interactions of two tie-variables and three outcome variables that are derived out of the partial dependence structures (see Section D of the Appendix) are depicted in Figure~\ref{fig:stats2}. Indirect contagion (Figure~\ref{fig:stats2}(e)) reflects processes where someone for example is likely to believe in something if the friends of their friends believe in it. Partner activity (Figure~\ref{fig:stats2}(f)) reflects processes where someone for example is likely to believe in something if the friends that believe in the same things are also very popular.

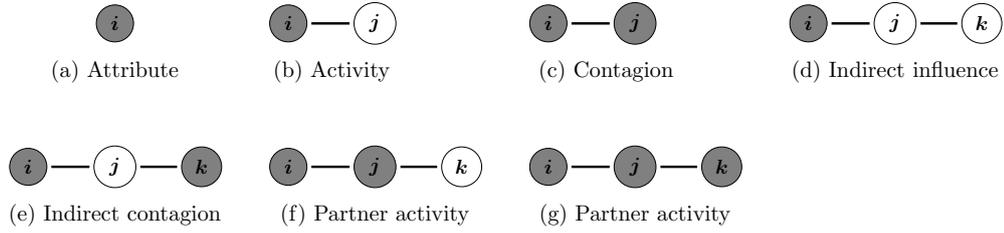
\begin{figure}

\begin{center}
\resizebox{380pt}{!}{
\begin{tikzpicture}
\begin{scope}
\boldmath \tikzstyle{every
node}=[x=3ex,y=2.5ex,shape=circle,minimum size=2ex]
\tikzstyle{every path}=[very thick, -stealth', shorten <=2pt,
shorten >=2pt]

%%% === a === %%%%
\node (a) at (3,10) {(a)  Attribute};
\node[draw,thin,fill=gray] (ai) at (3,12) {{\small $i$}} ; 
%%% === b === %%%%
\node (b) at (10.5,10) {(b)  Activity};
\node[draw,thin,fill=gray] (bi) at (9,12) {{\small $i$}} ;
\node[draw,thin] (bj) at (12,12) {{\small $j$}} ;
 \draw[-] (bi) -- (bj); 
%%% === c === %%%%
\node (c) at (20,10) {(c)  Contagion};
\node[draw,thin,fill=gray] (ci) at (18,12) {{\small $i$}} ;
\node[draw,thin,fill=gray] (cj) at (21,12) {{\small $j$}} ;
 \draw[-] (ci) -- (cj); 

%%% === d === %%%%
\node (d) at (30,10) {(d)  Indirect influence};
\node[draw,thin,fill=gray] (di) at (27,12) {{\small $i$}} ;
\node[draw,thin] (dj) at (30,12) {{\small $j$}} ;
\node[draw,thin] (dk) at (33,12) {{\small $k$}} ;
 \draw[-] (di) -- (dj);  \draw[-] (dk) -- (dj);

%%% === e === %%%%
\node (e) at (3,4) {(e)  Indirect contagion};
\node[draw,thin,fill=gray] (ei) at (0,6) {{\small $i$}} ;
\node[draw,thin] (ej) at (3,6) {{\small $j$}} ;
\node[draw,thin,fill=gray] (ek) at (6,6) {{\small $k$}} ;
 \draw[-] (ei) -- (ej);  \draw[-] (ek) -- (ej); 

%%% === f === %%%%
\node (f) at (12,4) {(f)  Partner activity};
\node[draw,thin,fill=gray] (fi) at (9,6) {{\small $i$}} ;
\node[draw,thin,fill=gray] (fj) at (12,6) {{\small $j$}} ;
\node[draw,thin] (fk) at (15,6) {{\small $k$}} ;
 \draw[-] (fi) -- (fj);  \draw[-] (fk) -- (fj); 

%%% === g === %%%%
\node (g) at (21,4) {(g)  Partner activity};
\node[draw,thin,fill=gray] (gi) at (18,6) {{\small $i$}} ;
\node[draw,thin,fill=gray] (gj) at (21,6) {{\small $j$}} ;
\node[draw,thin,fill=gray] (gk) at (24,6) {{\small $k$}} ;
 \draw[-] (gi) -- (gj);  \draw[-] (gk) -- (gj);

\end{scope}
\end{tikzpicture}}\\
\end{center}
\caption{Statistics associated with the indirect structural dependence assumption and the indirect dependent attribute assumption. Filled nodes indicate $y_i=1$, and unfilled nodes represent $y_i=0$ or $y_i=1$}\label{fig:stats2}
\end{figure}

\section{Inference}
% Section 3.2 Simulation
\subsection{Simulating from the model}
The expression of Eq.~\ref{eq:alaam} cannot be evaluated analytically as the normalising constant $\kappa(\theta)$ is a sum over all of $\mathcal{Y}$. Simulation for Markov random fields is however straightforward and has a long history. Many algorithms have been proposed and they typically draw on the conditional independence that implies that
\[
\text{logit}\Bigg \{ {\Pr}_{\theta}(Y_{i}=1|Y_{-i}=y_{-i},X=x) \Bigg \} = \theta^{\top} \{z(\Delta^{+}_{i}y,x)-z(\Delta^{-}_{i}y,x)\}
\]
where $z(\cdot)$ is the vector of sufficient statistics, $\Delta^{+}_{i}y$ is the vector $y$ with element $i$ set to one, and $\Delta^{-}_{i}y$ is the vector $y$ with element $i$ set to zero.  For a nearest neighbour algorithm we can update $y$ iteratively by selecting $i$ at random from $V$ and either update it using a Gibbs-update, or through a Metropolis updating step by proposing to change the value from $y_i$ to $1-y_i$. It is also possible to update multiple variables in parallel using Besag's (1974) coding scheme approach for the Ising mode. For example, outcomes for nodes that are isolated in a graph can be updated independently of all other values. Blocks of variables may also be updated independently of each other if they are well separated in the sense of Pattison et al. (2013).

Snijders (2002) discuss a number of sampling schemes for improving mixing for ERGMs and Butts (2018) has shown that perfect sampling is possible for ERGMs and, by implication, ALAAMs. Here we prefer the standard Metropolis algorithm since it is robust and has low memory overhead. Snijders (2002) developed the rule that the burnin for sampling from the ERGM should be a multiple of $n^2d(1-d)$, with the rationale that each variable is given an opportunity to change but the more ties there are (in the distribution), the longer the burning needs to be. The tie-no-tie algorithm, used for example by Caimo and Friel (2013), addresses the latter problem by having different proposal probabilities for null-ties and ties. Here we set the burnin to $\gamma n$ which, for same the multiplication factor $\gamma$, allows for longer burnin than that of Snijders (2002).

% 4.2 estimation
\subsection{Estimation\label{estimation}}
The main obstacle to Bayesian inference for the model of Eq.~\ref{eq:alaam}, is that the posterior is doubly intractable in the sense that both the normalising constant of the posterior and the likelihood are intractable. With prior distribution $\pi(\theta)$, the posterior distribution is
\[
\pi(\theta|y,x) = \frac{ \exp \{ \theta^{\top} z(y,x) - \psi(\theta)  \} \pi(\theta)} {\int \exp \{ \theta^{\top} z(y,x) - \psi(\theta) \} \pi(\theta)  \mathrm{d}\theta}{\text{,}}
\]
where we note that the numerator contains the intractable normalising constant $\psi(\theta)$ and the denominator involves an intractable integral (of an intractable expression). %MCMC may be performed by numerically approximating $\psi(\theta)$, either by interpolation on a grid (which is much more straightforward for auto logistic models than for ALAAM) or through simulation-based methods.
The auxiliary variable MCMC elegantly avoids having to evaluate  $\psi(\theta)$ by drawing variables from an auxiliary distribution with the same set of parameters  (M{\o}ller et al., 2006). The performance of the auxiliary-variable MCMC relies critically on how well the values of the parameters in the auxiliary distribution represent the true but unknown posterior.  The linked importance sample auxiliary variable MCMC (Koskinen, 2008) alleviates this issue by introducing bridging distributions, linking the reference distribution to the likelihood. The performance of the linked importance sample auxiliary variable MCMC is however dependent on a good choice of auxiliary variable parameters. The exchange algorithm (Murray et al., 2006) removes the need for the parameters of the auxiliary variable to be fixed and in the process not only reduces computational overheads but also automatically tunes the auxiliary parameters in the course of the MCMC. Caimo and Friel (2011) adopted the exchange algorithm to ERGMs approximating the Gibbs updating step by a Metropolis MCMC. Here we adopt this approximate exchange algorithm to ALAAM.

For the ALAAM, the exchange MCMC has as its target distribution the joint distribution
\[
\pi(\theta,y^{\ast},\theta^{\ast}|y,x,) \propto \exp \{ \theta^{\top} z(y,x) - \psi(\theta)  \} \pi(\theta) h(\theta^{\ast}|\theta) \exp \{ \theta^{\ast \top} z(y^{\ast},x) - \psi(\theta^{\ast})  \} 
\]
where $y^{\ast}$ is a variable with the same distributional form as $y$ but with a parameter $\theta^{\ast}$, the prior of which is $ h(\theta^{\ast}|\theta) $. In other words, $y^{\ast}$ follows the same ALAAM as data $y$ but with a different parameter.  Marginalising this joint posterior with respect to  $y^{\ast}$  and $\theta^{\ast}$, we obtain our desired posterior for $\theta$ given $y$.

A sample $\{ \theta_t,y_t^{\ast} ,\theta_t^{\ast} \}_{t=1}^T$, is generated through a two-step updating procedure in each iteration $t$. Given the current values $\theta_{t-1}$, $y_{t-1}^{\ast}$, and $\theta_{t-1}^{\ast}  $, we first propose $\theta^{\ast}$ from $ h(\theta^{\ast}|\theta_{t-1}) $, and conditional on the proposed value, draw $y^{\ast}$ from $p_{\theta^{\ast}}(\cdot|x) $. Given these proposed values, we propose to swap the parameters with probability $\min\{1,H\}$
\[
H= \frac{  p_{\theta^{\ast}}( y |x) p_{\theta_{t-1}}( y^{\ast} |x)   }      {   p_{\theta_{t-1}}( y |x)   p_{\theta^{\ast}}( y^{\ast} |x)  } \frac{ h(\theta_{t-1}|\theta^{\ast} ) }{ h(\theta^{\ast}|\theta_{t-1}) } \frac{\pi( \theta^{\ast} )  }{ \pi (\theta_{t-1}) }{\text{,}}
\]
setting $\theta_t=\theta^{\ast}$ and $\theta^{\ast}_{t}=\theta_{t-1}$. Similar to the case of (non-curved) ERGMs (Caimo and Friel, 2011), the acceptance probability $\alpha(  \theta_{t-1},\theta^{\ast} | y^{\ast}  )$ simplifies to 
%\[
%H=\exp\{ (\theta_{t-1} - \theta^{\ast})^{\top} ( z( y^{\ast} ,x) - z(y,x) ) \} \pi( \theta^{\ast} ) /  \pi (\theta_{t-1}) 
%\]
\begin{equation}
\alpha(  \theta,\theta^{\ast} | y^{\ast}  ) =  \min \Big[  1 ,  \exp\{   ( \theta - \theta^{\ast}   )^{\top}( z( y^{\ast},x  )  - z(y,x))    \}\pi( \theta^{\ast} ) /  \pi (\theta)   \Big].
\label{eq:hastingsratio}
\end{equation}
It is convenient to use a symmetric proposal distribution for $h(\cdot | \cdot )$. In particular, we propose a simplistic multivariate normal with mean vector $\theta_{t-1}$ and a variance-covariance matrix that is set to $cp^{-1/2}$ times the inverse of $Cov_{\theta_0}(z(Y,x),z(Y,x))$, approximated from a short initial sample $(y^{(t)})$ from the model defined by the initial value $\theta_0$. For exponential family models in canonical form we have that $I(\theta)=Cov_{\theta_0}(z(Y,x),z(Y,x))^{-1}$ (this procedure was also used by Koskinen et al., 2013, for tuning the algorithm for ERGM).
% clarification of proposal variance covariance
We set $\theta_0$ to the MLE under a model where contagion parameters are set to zero. As this model is equivalent to logistic regression, the MLE is readily available using standard estimation techniques.

% add reference to Everitt
We draw $y^{\ast}$ as described in Section 3.2 and hence the algorithm is an \emph{approximate} exchange sampler. Everitt (2012) discusses the implications of the approximation for the properties of the sampler, but ultimately the performance of the algorithm will depend on how well the draws of $y^{\ast}$ mixes for different $\theta$, something which will have to be decided on a case by case basis. 

\subsection{Missing data\label{sec:missing}}
Assume that we observe data $y$ only for a subset of actors given by the missing data indicator $(I_i)_{i=1}^n$, where $I_i=1$ if the response $y_i$ is unobserved for $i$ and $I_i=0$ if the response $y_i$ is observed for $i$. Following Rubin (1976) and Little and Rubin (1987) we define a missing data mechanism $f(I|y,\phi)$ conditional on the response variables where the parameter $\phi$ is distinct from the model parameters $\theta$. Initialising $y$ by assigning initial values to missing entries, with a prior $\pi(\phi)$, the estimation is carried out as above with two additional updating steps in each iteration. The first consists of updating the missing values and is done by, for each $i\in\{i\in V:I_i=1\}$, proposing to set $y_i=1-y_i$, and accepting this with probability
\[
\min \Big[   1, \exp\{\theta^{\top} (z(\Delta_{i}y,x)-z(y,x))\}  \frac{f(I|\Delta_{i}y,\phi)}{f(I|y,\phi)}  \Big]
\]
where $\Delta_{i}y$ is $y$ with element $i$ toggled and set to $1-y_i$. To update $\phi$, propose a move to $\phi^{\ast}$ drawn from a proposal distribution $q(\cdot|\phi)$, and accept this with probability
\[
\min \Big\{   1, \frac{f(I|y,\phi^{\ast})  \pi(\phi^{\ast}) }{   f(I|y,\phi)  \pi(\phi)}  \Big\}.
\]
If data are missing not at random (MNAR) we can define a missing data generating mechanism to test the sensitivity of our inference for $\theta$ to deviations from data being missing at random (MAR).

%When analysing social selection using ERGM, there are a number of ways of accounting for missing tie-variables (Handcock and Gile, 2010; Koskinen et al., 2010). When modelling tie-variables, accounting for key missing attributes is more complicated. While methods for jointly handling missing ties and missing attributes have been proposed (Koskinen et al., 2013), a more straightforward (albeit less principled) approach would be to perform multiple imputation for binary attributes using the ALAAM procedure above and, for example, using the network effects model (Marsden and Freidkin, 1994; Lenders, 2002; implemented in the routine {\texttt{lnam}} in the R-package {\texttt{sna}}, Butts, 2016) for continuous attributes.

\subsection{Goodness of fit}
For ERGMs it has become standard practice to evaluate model fit by considering the predictive distributions for a range of different features of the network (Hunter et al., 2008; Robins and Lusher, 2013). This is partly because the high-dimensional network space admits a number of projections. The outcomes in an influence model have a range-space $\mathcal{Y}$ that is considerably more straightforward to summarise. Given the suite of different statistics that the different dependence assumptions of Daraganova (2009) imply, it is still however necessary to consider a number of functions of $\mathcal{Y}\times \mathcal{X}$ as these may inform us of dependencies in data that we have not captured. Similar to the Bayesian goodness-of-fit (GOF) for ERGMs (Koskinen et al., 2010; Koskinen et al., 2013), the GOF distribution is the posterior predictive distribution, marginalised over the parameters.
%(and not based on a plug-in  point estimate, as in MPNet for ALAAM, Wang et al., 2014)
The predictive distribution $\{ y^{(t)}\}$ is obtained from drawing $y^{(t)}$ from the ALAAM defined by the posterior draw $\theta_t$. In the MCMC that generates the posterior draws, whenever $\theta$ is updated we set $\theta=\theta^{\ast}$. As the auxiliary variable $y^{\ast}$ is drawn from the distribution $p_{\theta^{\ast}}(\cdot|x)=p_{\theta}(\cdot|x)$, the draw of $y^{\ast}$ is also a draw from the posterior predictive distribution. Thus, if we let $y^{(t)}=y^{\ast}$ for every $t$ such that $\theta_t \neq \theta_{t-1}$, and  $y^{(t)}=y^{(t-1)}$, otherwise, we have a draw $\{ y^{(t)} \}$ from the posterior predictive distribution at the termination of the estimation algorithm. Note that the Bayesian GOF is based on draws of replicate data from the predictive distribution $\pi(\cdot | y )$ and as such accounts for uncertainty in parameters. Assuming that we have models $M_1,\ldots,M_K$ with posterior distributions $\pi(M_k|y)$, we can average the predictive distributions over models.
%(obviously, a GOF distribution is also available immediately from Phase 3 in MPNet and ).

\subsection{Model selection\label{sec:modelsel}}
Caimo and Friel (2013) propose an across-model procedure to evaluate model evidence for ERGM. They note that within-model estimation of evidence that relies on density estimation of the posterior breaks down for high-dimensional parameter vectors (greater than 5). Friel (2013) proposes an elegant method for estimation of Bayes factors of pair-wise nested models based on the MCMC updating in the exchange algorithm and demonstrates their application to two simple Markov random field models (the Ising model and a Markov two-star ERGM) (Everitt et al., 2017, propose direct estimation of the marginal likelihood using an importance sampling scheme that circumvents the need to evaluate $\psi(\cdot)$ using the trick of M{\o}ller et al., 2006).  Here we aim to provide a within-model estimation scheme that works for the types of complex models that you would expect when modelling outcomes in the social and behavioural sciences.  We follow an adoption of Chib and Jeliazkov (2001) that has previously been used for ERGMs (Koskinen, 2004). First we note from the so-called basic marginal likelihood identity that
\[
m(y) = \frac{ p_{\theta}(y|x)\pi (\theta) } { \pi(\theta | y )}
\]
where $m(y)$ is the marginal likelihood or equivalently the normalising constant of the posterior distribution of $\theta$ given $y$. This equality holds for any choice of $\theta$ and thus we can calculate the marginal likelihood by picking any value $\theta^{\prime}$ and evaluate the basic marginal likelihood for $\theta = \theta^{\prime}$.
We can use the path sampler to evaluate the likelihood ordinate (as in Hunter and Handcock, 2006, and  Caimo and Friel, 2013; for details see e.g. Gelman and Meng, 1998) but obtaining a good numerical approximation of the posterior ordinate $ \pi( \theta^{\prime} | y )$ is hard.

Proceeding by the method of Chib and Jeliazkov (2001), we define the subkernel of the M-H update as
\[
p(\theta,\theta^{\ast} , y^{\ast}) = \alpha(  \theta,\theta^{\ast} | y^{\ast}  ) h( \theta^{\ast}  \mid \theta )p_{\theta^{\ast}}(y^{\ast} \mid x),
\]
in which $\alpha(  \theta,\theta^{\ast} | y^{\ast}  )$ is defined as (\ref{eq:hastingsratio}).
By construction
\begin{eqnarray}
\label{eq:detbal}
\pi(\theta \mid y ) p(\theta,\theta^{\ast} , y^{\ast}) &=&\pi( \theta^{\ast} \mid y ) p(\theta^{\ast} ,\theta , y^{\ast})\\
\pi( \theta \mid y )p_{\theta^{\ast}}(y^{\ast} \mid x)\alpha(  \theta,\theta^{\ast} | y^{\ast}  ) h( \theta^{\ast}  \mid \theta ) &=& \pi( \theta^{\ast} \mid y )p_{\theta}(y^{\ast} \mid x)\alpha( \theta^{\ast} , \theta| y^{\ast}  ) h( \theta  \mid \theta^{\ast} ) \nonumber,
\end{eqnarray}
which can be verified by direct calculation. Upon integrating both sides of (\ref{eq:detbal}) with respect to  $\theta$ over $\Theta$ and taking the sum over $y^{\ast}$ we obtain {\small{
\[
\int \Big[ \sum_{y^{\ast} \in \mathcal{Y}} p_{\theta^{\ast}}(y^{\ast} \mid x)\alpha(  \theta,\theta^{\ast} | y^{\ast}  )  h( \theta^{\ast}  \mid \theta ) \Big] \pi( \theta \mid y ) \mathrm{d}\theta= \pi( \theta^{\ast} \mid y )\int\Big[ \sum_{y^{\ast} \in \mathcal{Y}} p_{\theta}(y^{\ast} \mid x)\alpha( \theta^{\ast} , \theta| y^{\ast} )\Big]  h( \theta  \mid \theta^{\ast} ) \mathrm{d}\theta.
\]}}
We recognise the summands as expectations with respect to the likelihood, and the integrands as expectations with respect to the posterior (LHS) and the proposal distribution (RHS). Consequently, solving for $\pi( \theta \mid y )$, for any parameter value $\theta^{\prime}$, the posterior ordinate can be written as a ratio of expectations
\begin{equation}
\pi( \theta^{\prime} \mid y ) =
 \frac{E_{\pi(\theta \mid y )} \Big[ E_{p_{\theta^{\prime}}(y^{\ast} \mid x)} \Big\{ \alpha(  \theta,\theta^{\prime} | y^{\ast}  )  h( \theta^{\prime}  \mid \theta ) \Big\} \Big] }
 { E_{h(\theta \mid \theta^{\prime})} \Big[ E_{p_{\theta}(y^{\ast} \mid x)} \Big\{ \alpha( \theta^{\prime} , \theta | y^{\ast} ) \Big\} \Big] }
\label{eq:postord}
\end{equation}

We can evaluate the numerator using the Monte-Carlo estimate, taking $\theta$ from our posterior draws $\{ \theta_t \}_{t=1}^T$ %as
%\[
%\int_{\theta^{\ast}} \sum_{y^{\ast}}\pi(\theta|y,x,y^{\ast},\theta^{\ast}) {\mathrm{d}}\theta^{\ast} = \pi(\theta|y,x)
%\]
and for the inner expectation we can take a sample of auxiliary variables for each $  \theta^{\prime}$. We may also change the order of the expectations in the numerator, meaning that we draw one large sample $\{  y^{(g)} \}_{g=1}^G$ from the distribution defined by $\theta=\theta^{\prime}$ and average $ \alpha (  \theta^{(t)}, \theta^{\prime} | y^{(g)}  ) h(\theta^{\prime}  | \theta^{(t)} )$. For the denominator we draw a number of $\{ \theta_j \}$ from the proposal distribution $ h(\cdot | \theta^{\prime} ) $ and similarly calculate Monte Carlo averages of the conditional acceptance probability $\alpha({ \cdot})$ across samples from $\mathcal{Y}$.

With missing data in $Y$, the likelihood is given by $\exp\{\psi(\theta,\phi,I)-\psi(\theta)\}$ where $\psi(\theta,\phi,I)=\log [\sum_{y_i:I_i=1}\exp\{ \theta^{\top}z(y,x) \}  f(I|y,\phi)]$. We can evaluate $\psi(\theta,\phi,I)$ using the path sampler with the restriction that $Y_i=y_i$ are fixed for $i$ such that $I_i=0$. With missing data Eq. (\ref{eq:postord}) needs to be modified to account for the uncertainty in the missing outcomes. When evaluating the acceptance probability in the numerator of Eq. (\ref{eq:postord}), we may take the Monte Carlo average over the joint posterior of $\theta$ and $\{y_i|I_i=1\}$ for the corresponding draws from the joint posterior. For the denominator  of Eq. (\ref{eq:postord}), the Monte Carlo average will be taken with respect to draws from $h(\theta | \theta^{\prime} )$ and draws of $\{y_i|I_i=1\}$ from the conditional distribution $p_{ \theta^{\prime}}(y| x)$. The Monte Carlo estimate of (\ref{eq:postord}) with missing data is written as follows
\begin{equation}
\hat{\pi}(\theta^{\prime}|\{ y_i:I_i=0\} )= \frac{ \sum_{ g=1}^G G^{-1}  \sum_{t=1}^T  T^{-1} \alpha(\theta_t,\theta^{\prime} | y^{(g)}, v_t) h(\theta^{\prime} | \theta_t) }  {    \sum_{j=1}^J J^{-1}  \sum_{m=1}^M M^{-1}  \alpha(\theta^{\prime},\theta_j | y^{(j,m)}, u_j) }{\text{,}}
\label{eq:postest}
\end{equation}
where $\{v_t \}$ are posterior draws of $\{ y_i:I_i=1\}$, $\{ y^{(g)}\}$ are draws from $p_{ \theta^{\prime}}(\cdot |x)$, $\{\theta_j \}$ are independent draws from $h(\cdot | \theta^{\prime})$, $\{u_j \}$ are draws of $\{ y_i:I_i=1\}$ from the model $p_{ \theta_j}(\cdot | x )$, and finally $\{ y^{(j,m)} \}$ are $M$ draws from $p(\cdot | \theta_j)$. The numerator in Eq.~\ref{eq:postest} is computationally cheap to evaluate as we only need one large sample $\{y^{(g)}\}$ from the distribution defined by $\theta^{\prime}$. The denominator in Eq.~\ref{eq:postest} does however require a sample of size $M$ from the model defined by $\theta_j$ for all $j=1,\ldots,J$. The variance of the estimator is not very sensitive to the size of $M$ and $G$ and setting both to about 100 appears sufficient. As discussed in Chib and Jeliazkov (2001), the estimator requires a well mixing sample from the posterior for the numerator of a length $T$ of the order 10,000 to 20,000. Here we have to factor in the variation in the evaluation of the likelihood (as well as the acceptance probability) and a precise estimate is likely to require $T$ in excess of 20,000. (A brief illustration of the effect of different sample sizes $T$ on the estimator is provided in Section F of the Appendix.)

\subsubsection{Prior distributions\label{sec:priors}}
There are good reasons for performing inference for ALAAMs with prior distributions that are proper. With an improper prior distribution for $\theta$, the posterior distribution is proper if the observed vector of statistics is in the relative interior of the convex hull on $\mathcal{Z}$, where  $\mathcal{Z}$ is the image of $\mathcal{Y}$ under $z(\cdot)$%(this follows trivially from the properties of exponential families, Barndorff-Nielsen,1978; see e.g. Koskinen et al., 2010)
. Since there are instances where $z(y,x)$ does not fall in the (relative interior of the) convex hull on $\mathcal{Z}$ (Handcock , 2003), a proper prior distribution formally is a safeguard against the risk of the posterior not being defined.
The Bayes factor $\pi(y|M_i)/\pi(y | M_j)$ is only properly defined if the prior distributions for the parameters of both models are proper. A convenient choice for prior distribution for the canonical parameters for an exponential family model is a multivariate normal distribution $N_p(\mu,\Sigma)$. While one can be motivate setting $\mu=0$ a priori to reflect no bias on the parameters, setting the scale through $\Sigma$ is less straightforward. For related binomial models, Chen et al. (2008) argue the merits of using Jeffreys' prior (Jeffreys, 1946). Here, this would translate to the prior being $\pi(\theta)\propto | I(\mu_0)|^{1/2}$, for $\mu=0$ which motivates the scalable normal prior with variance covariance matrix $\lambda I(\mu_0)^{-1}$. The information matrix is straightforward to obtain as the Monte Carlo estimate of the variance covariance of the model sufficient statistics under $\theta=\mu_0$. For some data sets where $\sum_iy_i$ is small, one can motivate using a data-dependent prior with  $(\mu_0)_1=\hat{\theta}_{MLE}$, where $\hat{\theta}_{MLE}=- \log(\bar{y}-1)$. As $\lambda$ will shrink the prior distribution and pull parameters towards the origin, setting $(\mu_0)_1=\hat{\theta}_{MLE}$ will reduce the influence of the intercept which is largely a nuisance.

\subsubsection{Posterior deviance\label{sec:postdev}}
To evaluate model fit with constant or reference priors, posterior predictive p-values (Meng, 1994) may be applied for any function of the network and attributes that are typically used in GOF (Hunter et al., 2008; Robins and Lusher, 2013). For single value summaries of model fit we may also consider functions of the deviance.  In the context of complex network models, Aitkin et al. (2017) considered evidence in terms of the posterior distribution of the deviance (a full discussion of this approach is given in Aitkin, 2010). This provides a useful graphical representation of relative fit of a model that can be summarised using the deviance information criterion (Spiegelhalter et al., 2002; Gelman et al., 2004). As the likelihood of Eq.~\ref{eq:alaam} is intractable, we need to evaluate the log-likelihood for each draw $\theta_t$ numerically using the path-sampler (Hunter and Handcock, 2006; Gelman and Meng, 1998). With missing data defined as in Section~\ref{sec:missing}, the likelihood is estimated as $\hat{\ell}(\theta_t)=\hat{\psi}(\theta_t,\phi,I)-\hat{\psi}(\theta_t)$. Here the likelihood is estimated using the path sampler relative to the MLE for a nested independent model. 

\section{Applications\label{sec:applications}}
% possibility of using standard dataset but most standard datasets not amenable to influence model
We demonstrate the proposed inference procedures using three data sets, two sociocentric school networks and one snowball-sampled data set on unemployment status.
%The first dataset explores the dependence of attitudes of individuals on their peers and is of a canonical network form, with a network census of a single school class, something which allows us to illustrate the basic features of the modelling framework. The second dataset has several features that you might expect to come across in studies of substance use, attitudes, and behaviours in adolescents and children. We illustrate how to perform the inference procedure in the face of missing data. The dataset also has outcomes for pupils across several school classes, something which allows us to explore how the strength of social influence may be tested across settings. Finally, we adopt the Bayesian inference scheme to snowball-sampled data. 
\subsection{Masculine attitudes in a school class}
\begin{figure}

\centering
    \makebox{
\includegraphics[scale=.6]{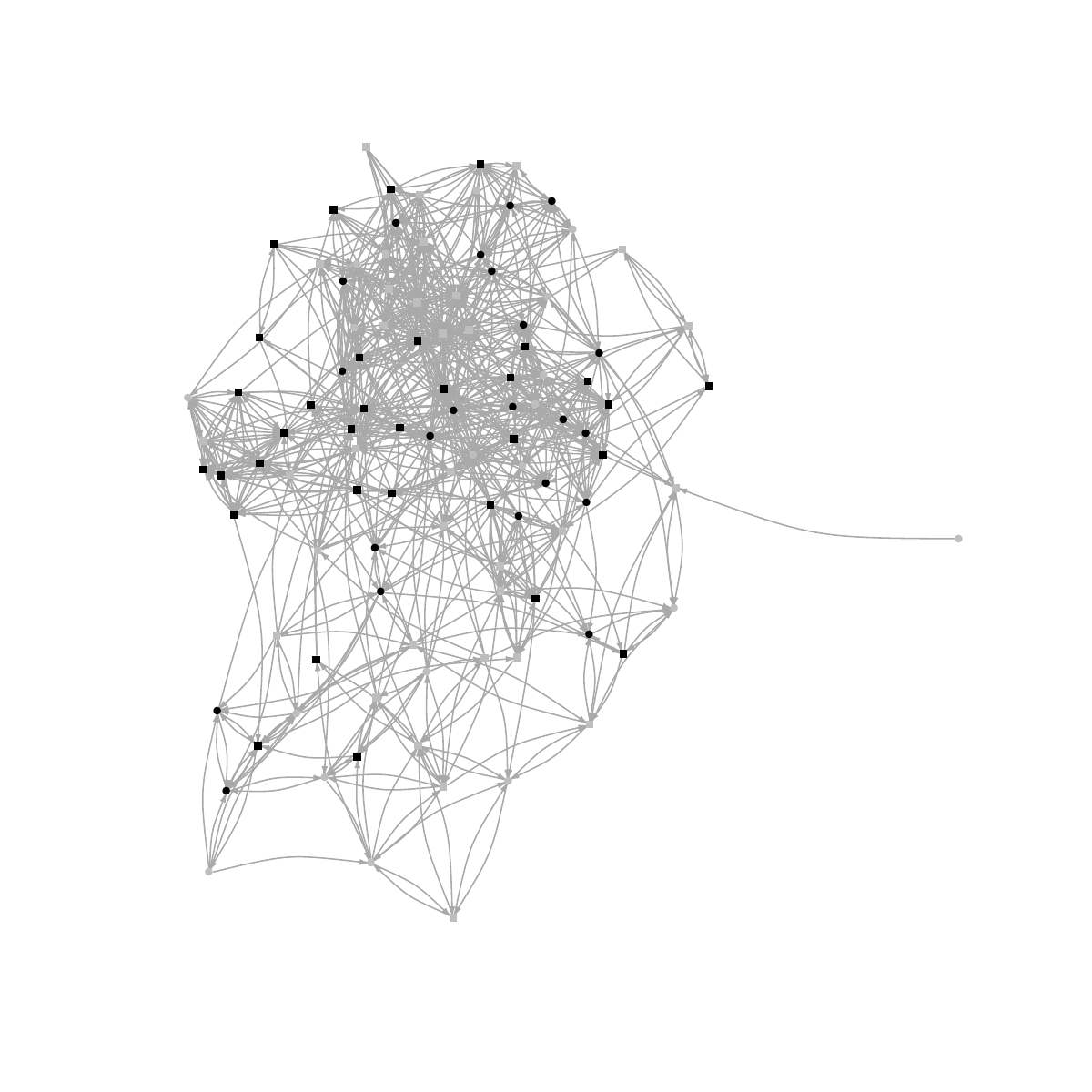}}

\caption{\label{fig:deansociogram} Friendship network among 106 pupils in an all-male school. Dominant culture indicated by squares (1) and circles (0), and outcome black ($y_i=1$), grey ($y_i=0$), for high and low masculinity index, respectively}
\end{figure}

Lusher and Dudgeon (2007) %(see also Lusher and Robins, 2009) 
developed a scale, MAI, for measuring male dominance attitudes. In school classes it may be of interest to know if a (male) pupil's attitudes to masculinity is contingent on those of his friends. MAI scores as well as friendship nominations were collected for 106 pupils in a Year 10 level in a single-sex, religious secondary school in Australia (Lusher, 2011). The response variable $y_i$ is the MAI dichotomised at the mean. Controls are: `dominant culture'  (indicates if $i$ has an Anglo-Australian ethno-cultural background (1) or not (0)); the socio-economic status of the pupil's household (as measured by standardised SES based on postcode); the occupational score for the father of the pupil%(original range is 0 to 100 according to Jones and McMillan, 2001, but here standardised)
; the equivalently defined occupational score for the mother of the pupil (see Lusher, 2011, for further details of the network data).

\subsubsection{Direct contagion}
\begin{figure}
\centering
    \makebox{
\includegraphics[scale=0.65]{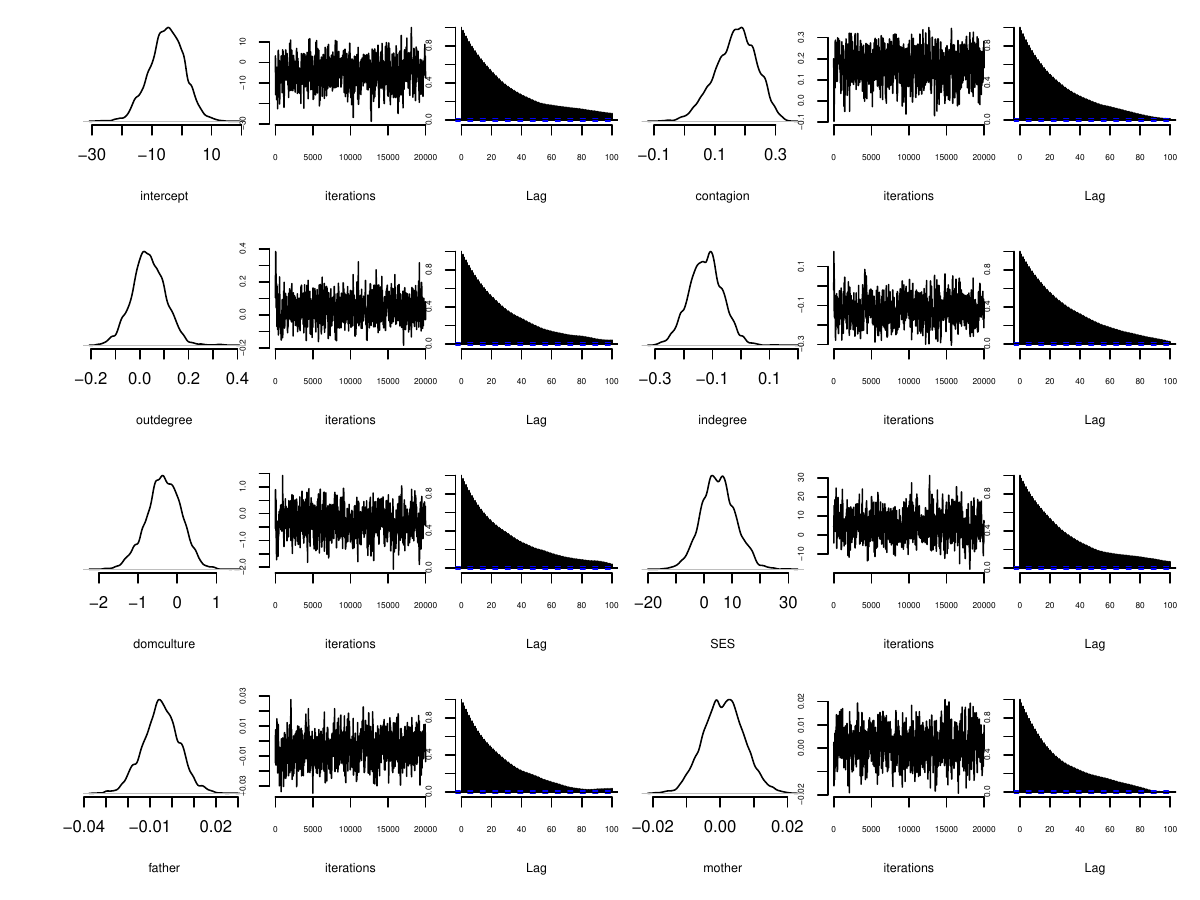}}
\caption{\label{fig:dean106MCMC}MCMC output for ALAAM contagion-model for masculine attitudes in an Australian school. Density estimate, trace plot, and SACF for each of the parameters reported in Table~\ref{tab:dean106}}
\end{figure}

Figure~\ref{fig:dean106MCMC} provides the MCMC output for a model under assumption [DC] with 20,000 draws using the standard settings of Section~\ref{estimation}, namely with the proposal variance-covariance matrix based on $cI(\tilde{\theta})$, where $I(\theta)$ is the inverse of the covariance matrix of the statistics under $\theta$; $\tilde{\theta}$ estimated from a simulation of statistics under the MLE for a logistic regression with contagion parameters set to $0$; and the tuning constant is set to $c=1$, and the multiplication factor for drawing from the likelihood is set to $\gamma = 30$. These settings will be used also for the rest of the examples unless otherwise specified. The auto-correlation for the contagion parameter is fairly large even at large lags. This can be improved upon by setting the proposal covariance matrix equal to the covariance of the posteriors (this reduces the sample autocorrelation function, SACF, greatly). According to the posterior summaries provided in Table~\ref{tab:dean106}, there is  evidence for a positive contagion parameter.

\begin{table}
\caption{\label{tab:dean106}Posterior summaries for an ALAAM and an LNAM of contagion of masculine attitudes in a school in Australia}
\centering
\fbox{%
\begin{tabular}{lrrrrrrrrr}
  \hline
 & \multicolumn{7}{c}{ALAAM}&  \multicolumn{2}{c}{LNAM}\\
 & mean & sd & ESS & SACF 10 & SACF 30 & 2.5 perc & 97.5 perc &  mean & sd \\ 
  \hline
intercept & -4.94 & 5.78 & 269.73 & 0.72 & 0.37 & -16.23 & 6.33 & -7.52 & 3.91 \\ 
  contagion & 0.17 & 0.07 & 251.07 & 0.70 & 0.35 & 0.03 & 0.29 && \\ 
  outdegree & 0.03 & 0.07 & 292.16 & 0.72 & 0.37 & -0.10 & 0.16& -0.04 & 0.03   \\ 
  indegree & -0.12 & 0.06 & 293.03 & 0.74 & 0.42 & -0.23 & -0.00& 0.13 & 0.04  \\ 
  domculture & -0.32 & 0.43 & 235.78 & 0.70 & 0.38 & -1.28 & 0.47 & -0.61 & 0.31  \\ 
  SES & 5.04 & 6.14 & 271.11 & 0.73 & 0.38 & -7.90 & 17.11  & 7.85 & 4.11  \\ 
  father & -0.00 & 0.01 & 281.46 & 0.68 & 0.34 & -0.02 & 0.01 & -0.01 & 0.01  \\ 
  mother & 0.00 & 0.01 & 285.92 & 0.68 & 0.31 & -0.01 & 0.01 & 0.01 & 0.00 \\ 
alpha & 		&		&		&		&		&		&	 &0.49 & 0.18 \\ 
   \hline
\end{tabular}}
\end{table}

\subsubsection{Indirect contagion}
To infer whether there is evidence of influence on MAI being transmitted through indirect ties under assumption [IDA], we add the statistic  $y_i\sum_{j}y_j \sum_{k,k\neq i,j}x_{ik}x_{kj}$%, corresponding to the configuration (c) in Figure~\ref{fig:morestats}
.  In addition we include a statistic for the number of nodes that are reachable from an individual, the number of indirect ties $\sum_i y_i \sum_{j,k}x_{ik}x_{kj}$. The potential for a brokerage effect on $Y$ is controlled for by the mixed 2-path effect $\sum_i y_i \sum_{j,k}x_{ij}x_{ik}$. The results for the elaborated model are provided in Table~\ref{tab:dean106ind}. The introduction of the additional contagion effect reduces the direct contagion (posterior correlation of $-.69$), making interpretation less conclusive than in the simpler model. The number of indicted ties is positive with a large posterior probability suggesting that pupils that are indirectly connected to many others are likely to have masculine attitudes.

\begin{table}
\caption{\label{tab:dean106ind}Posterior summaries for an ALAAM with indirect contagion of masculine attitudes in a school in Australia}
\centering
\fbox{%
\begin{tabular}{lrrrrrrr}
  \hline
 & mean & sd & ESS & SACF 10 & SACF 30 & 2.5 perc & 97.5 perc \\ 
  \hline
intercept & -6.83 & 6.53 & 314.11 & 0.78 & 0.44 & -20.14 & 5.31 \\ 
  contagion & 0.21 & 0.13 & 281.64 & 0.79 & 0.51 & -0.05 & 0.48 \\ 
  indirect cont & -0.02 & 0.02 & 275.48 & 0.79 & 0.52 & -0.06 & 0.01 \\ 
  outdegree & -0.48 & 0.27 & 242.09 & 0.80 & 0.54 & -1.02 & 0.02 \\ 
  indegree & 0.07 & 0.16 & 324.60 & 0.79 & 0.50 & -0.27 & 0.38 \\ 
  brokerage & -0.01 & 0.02 & 323.54 & 0.79 & 0.51 & -0.05 & 0.02 \\ 
  indirect ties & 0.07 & 0.03 & 211.01 & 0.80 & 0.53 & 0.02 & 0.13 \\ 
  domculture & -0.41 & 0.49 & 292.16 & 0.79 & 0.51 & -1.30 & 0.59 \\ 
  SES & 7.28 & 6.64 & 305.18 & 0.77 & 0.43 & -6.32 & 20.32 \\ 
  father & -0.01 & 0.01 & 1256.67 & 0.78 & 0.46 & -0.03 & 0.01 \\ 
  mother & -0.00 & 0.01 & 305.47 & 0.79 & 0.49 & -0.02 & 0.01 \\ 
   \hline
\end{tabular}}
\end{table}

%\noindent{\textbf{I have implemented the (a)-(c) of Figure 5 as well as (a) and (b) from Figure 6 and am currently fitting different models to see what works; I have not yet decided if i should fit higher order contagion effects without including all the lower order interactions}}

\subsubsection{GOF}
The posterior predictive distributions for some functions of $(Y,X)$ are provided in Table~\ref{tab:dean106GOF}. For reference, predictive distributions for a latent network effects model (LNAM) are provided (with estimates in Table~\ref{tab:dean106}). For this model we assume that there exists an $n\times1$ vector $u$ that follows the standard network effects model $u=\alpha Wu+\beta^{\top}B+\epsilon$, where $W$ is the row-normalised adjacency matrix, $\alpha \in (-1,1)$ is the network effects parameter (Marsden and Friedkin, 1994), $B$ is a matrix with the same fixed covariates as for the ALAAM in Table~\ref{tab:dean106}, and $\epsilon$ are i.i.d. standard normal variates. As $Y$ is binary, we use $u$ as the latent variable for a probit link-function by letting $y_i= \mathbbold{1}\{u_i \geq 0\}$. Estimation of $\alpha$ and $\beta$ largely follows Koskinen and Stenberg (2012). We assume the same form of prior, as described in Section~\ref{sec:modelsel}, for the $\beta$ in the network effects model and $\theta$ in the complex contagion model with the exception that the $\alpha$ in the former is not included in the regression parameters.

%\begin{table}
%\caption{\label{tab:dean106NAM}Posterior summaries for latent network autocorrelation model (LNAM) with controls estimated for contagion-model for masculine attitudes in a school in Australia}
%\centering
%\fbox{%
%\begin{tabular}{lrr}
%  \hline
% & mean & sd \\ 
%  \hline
%intercept & -7.52 & 3.91 \\ 
%  alpha & 0.49 & 0.18 \\ 
%  outdegree & -0.04 & 0.03  \\ 
%  indegree & 0.13 & 0.04 \\ 
%  domculture & -0.61 & 0.31  \\ 
%  SES & 7.85 & 4.11  \\ 
%  father & -0.01 & 0.01  \\ 
%  mother & 0.01 & 0.00  \\ 
%   \hline
%\end{tabular}}
%\end{table}

For this relatively limited set of attribute and network interactions, the ALAAM marginally outperforms the LNAM judging by the posterior predictive p-values of Table~\ref{tab:dean106GOF}. However, for this dataset there is no clear evidence of the LNAM completely failing to reproduce any of the observed statistics.
% but for larger datasets with higher-order interaction effects LNAM is likely to be inadequate. 
The goodness-of-fit does also illustrate that the simpler specification of the ALAAM is sufficient for explaining higher-order dependencies such as indirect contagion.

\begin{table}
\caption{\label{tab:dean106GOF} Posterior predictive p-values for ALAAM and latent network autocorrelation models of Table~\ref{tab:dean106}}
\centering
\fbox{%
\begin{tabular}{lrrrrr}
  \hline
   & & \multicolumn{2}{c}{ALAAM}&  \multicolumn{2}{c}{LNAM}\\
statistic & observed & mean & p-value & mean & p-value \\ 
  \hline
intercept & 55.00 & 55.87 & 0.21 & 53.02 & 0.20 \\ 
  direct contagion & 272.00 & 277.94 & 0.25 & 250.54 & 0.17 \\ 
  reciprochal contagion & 75.00 & 78.17 & 0.24 & 74.39 & 0.22 \\ 
  indirect contagion & 2069.00 & 2132.74 & 0.26 & 1880.53 & 0.17 \\ 
  closure contagion & 763.00 & 790.04 & 0.26 & 695.57 & 0.17 \\ 
  transitive contagion & 456.00 & 534.64 & 0.24 & 456.68 & 0.28 \\ 
  indegree & 428.00 & 432.53 & 0.24 & 393.71 & 0.14 \\ 
  outdegree & 478.00 & 488.87 & 0.21 & 481.77 & 0.23 \\ 
  two-paths & 3859.00 & 3972.61 & 0.22 & 3786.91 & 0.22 \\ 
  out-2-star & 2362.00 & 2404.59 & 0.22 & 2452.69 & 0.17 \\ 
  in-2-star & 1780.00 & 2082.39 & 0.19 & 1701.24 & 0.18 \\ 
  out-triangles & 1368.00 & 1404.57 & 0.21 & 1385.58 & 0.22 \\ 
  in-triangles & 1210.00 & 1191.38 & 0.23 & 1048.03 & 0.11 \\ 
  transitive triangles & 1004.00 & 1033.12 & 0.23 & 951.75 & 0.19 \\ 
  indirecct ties & 3957.00 & 3914.89 & 0.25 & 3895.86 & 0.24 \\  
   \hline
\end{tabular}}
\end{table}

% Section 5.2
\subsection{Stockholm Birth Cohort}

\begin{figure}
\centering
    \makebox{
\includegraphics[scale=.4]{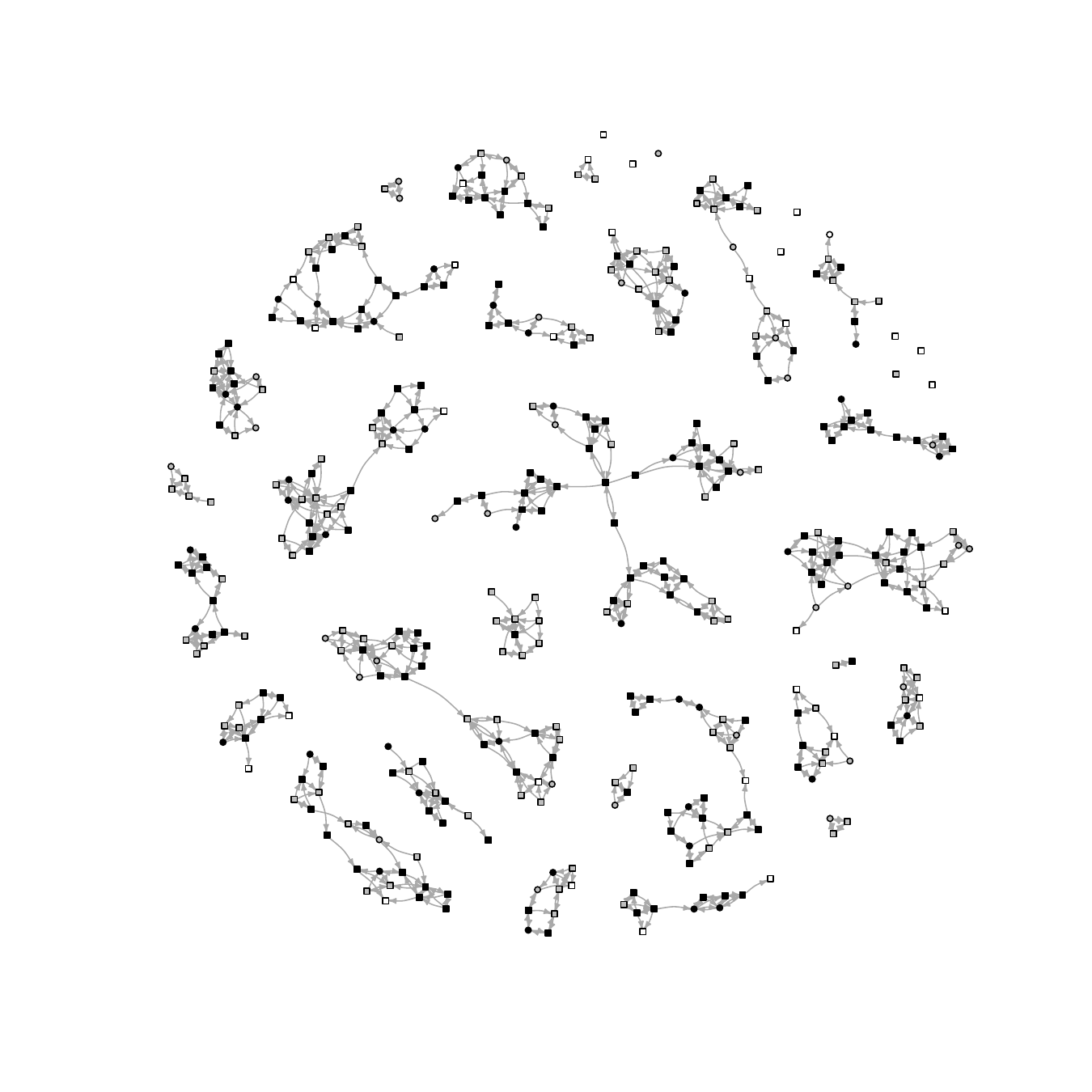}}
\caption{\label{fig:sbcsociogram} Best friend network in four schools in the Stockholm Birth Cohort. Sex indicated by squares (girl) and circles (boys), and outcome -- intention to proceed to higher secondary education -- black ($y_i=1$), grey ($y_i=0$), and white for missing}
\end{figure}

The Stockholm Birth Cohort is a large cohort study in the Stockholm Metropolitan area that includes detailed surveys and school-class network data (Stenberg and  V{\aa}ger{\"o}, 2006; %Stenberg et al. 2007; 
Stenberg, 2018). The networks are the best-friend nominations of school children and for each pupil there are a range of sociological, psychological, and educational variables. The survey was carried out in May 1966 when the pupils were nearing the end of the sixth grade. This is when they would have started considering whether they were going to proceed to higher secondary education (grades 10 and above) and been talking about this with their peers. We chose for our example 19 school classes out of the 1966 survey.
%, six of which are from a school in a suburb in the south of Stockholm and the rest are from three inner-city schools.
We let $X$ be the directed best-friend network (this had a cap of three nominations), and $y$ be indicators $y_i=1$ of whether pupils $i$ said that they intended to proceed to higher secondary school, and $y_i=0$ otherwise (in accordance with the model of Koskinen and Stenberg, 2012). By design there are no ties between pupils in different school classes. %The average network size is $21.2$, the average density is $0.14$ with a minimum of $0.084$ and a maximum of $0.45$ (the most dense network is the smallest network consisting of 5 pupils). 
The proportion of missing entries range from $0$ to $0.286$ with an average of $0.079$. We apply the ALAAM specified by assumption [DC] but set the parameter for out-stars (of the form $y_i x_{i,+}$) to zero as the nominations were capped at three and there is little variance in the out-degree distribution. In addition to this structural part, we control for: sex (female:1); family support (an 11-point scale measuring the family's attitude toward school ranging from least positive, 0 to most positive 10); average school marks (scaled to range from 0 to 10); an indicator of whether the father belongs to the top social class or not.%'s social class is one (the highest social class in a 5-point ordinal categorisation). 

The results from the MCMC with 10,000 iterations with constant priors $\pi(\theta)\propto c$ are summarised in Table~\ref{tab:resSBC} (the table is based on default settings with a burnin of 1000 and thinning of 20 iterations, and the same proposal as in the previous example; $\gamma = 7.5$). Mixing of the MCMC can be said to be satisfactory with default settings. There is strong evidence for a positive family attitude to school and high grades to increase the likelihood of the intention to proceed to higher education. The evidence is inconclusive for other effects. In particular, the contagion parameter is positive with posterior probability $0.93$.

%%% NB: strong effect of contagion might be amplified by between-class differences
%%% consider fixed effect for school

%\begin{figure}
%\centering
%    \makebox{
%\includegraphics[scale=0.7]{sbcNoInteractionp7.pdf}}
%\caption{\label{fig:sbcMCMC}MCMC output for contagion-model for progression to upper-secondary school in SBC}
%\end{figure}

\begin{table}
\caption{\label{tab:resSBC}Model 1 posterior summaries for contagion-model for progression to upper-secondary school in SBC (Posterior means, sd, and probability interval based on a thinned sample of 10,000 iterations, taking every 20th iteration, with burnin of 1000; SACF and ESS based on un-thinned sample) }
\centering
\fbox{%
\begin{tabular}{lrrrrrrr}
  \hline
 & mean & sd & ESS & SACF 10 & SACF 30 & 2.5 perc & 97.5 perc \\ 
  \hline
intercept & -9.67 & 1.11 & 178.03 & 0.68 & 0.32 & -11.83 & -7.51 \\ 
  contagion & 0.16 & 0.10 & 183.10 & 0.68 & 0.32 & -0.04 & 0.35 \\ 
  indegree & -0.07 & 0.11 & 183.55 & 0.67 & 0.32 & -0.29 & 0.13 \\ 
  sex & -0.09 & 0.29 & 134.35 & 0.70 & 0.39 & -0.66 & 0.47 \\ 
  family attitude & 0.48 & 0.09 & 164.22 & 0.70 & 0.32 & 0.33 & 0.65 \\ 
  marks & 0.99 & 0.15 & 168.66 & 0.68 & 0.32 & 0.69 & 1.28 \\ 
  social class 1 & 0.59 & 0.32 & 198.40 & 0.66 & 0.24 & -0.06 & 1.19 \\ 
   \hline
\end{tabular}}
\end{table}

\subsubsection{Testing difference in contagion}
The classes come from 4 schools that differ in socio economic status of uptake area as reflected in the composition of social class of pupils. We divide the schools into one subset with less than 15\% of students (across school classes) from the highest social class and a subset with more than 15\% of students from the highest social class. Table~\ref{tab:resSBCinteract} present the results for a model ($M_2$) that includes an interaction of the contagion parameter and an indicator for the type of school ($=1$ for schools with low proportion of pupils from the highest social class) as well as the main effect. There is stronger evidence than for model 1 for a contagion effect (the contagion parameter is positive with $0.988$ posterior probability). There is weak evidence for contagion being absent in schools with a lower proportion of pupils from the highest social class (the posterior distribution for $\theta_2+\theta_9$ has a mean of $0.0328$ and a standard deviation of $ 0.144$ and is negative with $ 0.435$ posterior probability).

\begin{table}
\caption{\label{tab:resSBCinteract}Model 2 posterior summaries for contagion-model for progression to upper-secondary school in SBC with an interaction between social contagion and social class (Posterior means, sd, and probability interval based on thinned sample of 10,000 iterations, taking every 20th iteration, with burnin of 1000; SACF and ESS based on un-thinned sample) }
\centering
\fbox{%
\begin{tabular}{lrrrrrrr}
  \hline
 & mean & sd & ESS & SACF 10 & SACF 30 & 2.5 perc & 97.5 perc \\ 
  \hline
intercept & -10.13 & 1.19 & 168.32 & 0.76 & 0.44 & -12.81 & -8.04 \\ 
  contagion & 0.24 & 0.12 & 143.31 & 0.72 & 0.39 & 0.02 & 0.48 \\ 
  indegree & -0.08 & 0.12 & 122.80 & 0.75 & 0.41 & -0.33 & 0.13 \\ 
  sex & -0.09 & 0.28 & 126.04 & 0.76 & 0.45 & -0.69 & 0.47 \\ 
  family attitude & 0.48 & 0.08 & 140.26 & 0.72 & 0.38 & 0.34 & 0.65 \\ 
  marks & 1.01 & 0.14 & 265.08 & 0.72 & 0.40 & 0.76 & 1.31 \\ 
  composition & 0.91 & 0.55 & 137.33 & 0.74 & 0.39 & -0.25 & 1.97 \\ 
  social class 1 & 0.57 & 0.34 & 143.59 & 0.73 & 0.37 & -0.07 & 1.21 \\ 
  contagion $\times$ social class 1 & -0.21 & 0.16 & 152.15 & 0.72 & 0.37 & -0.51 & 0.11 \\ 
   \hline
\end{tabular}}

\end{table}

%\begin{figure}
%\centering
%    \makebox{
%\includegraphics[scale=0.7]{contagionSBCinteraction.pdf}}
%\caption{\label{fig:postdensSBCinteraction}The posteriors of contagion parameters for model 1 and model 2 and the interaction of contagion and composition with 95\% probability intervals}
%\end{figure}

\begin{figure}
\centering
    \makebox{
\includegraphics[scale=.75]{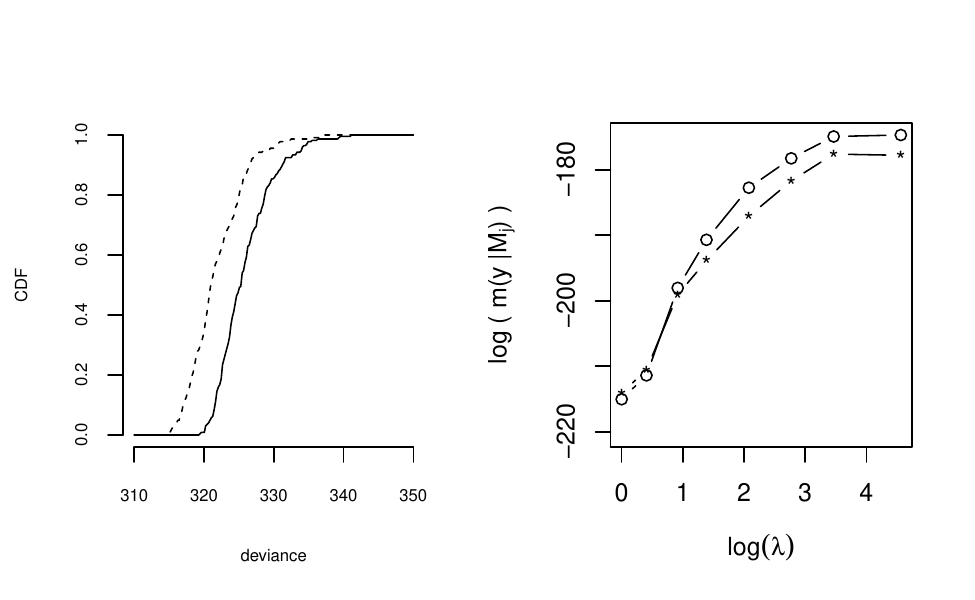}}
\caption{\label{fig:aitkinplot}Left panel: The posterior deviance for model 1 (Table~\ref{tab:resSBC}) (solid) and model 2 (Table~\ref{tab:resSBCinteract}) (dashed) . Right panel, model evidence for model 1 ($o-$) and model 2 ($*-$) for prior $N_p(\mu_0,\lambda I(\mu_0)^{-1})$ with scale $\lambda$ and $\mu_0=0$}
\end{figure}

%There is an indication of a difference in the posteriors in Figure~\ref{fig:postdensSBCinteraction} but can we test $M_1$ against $M_2$? 
Consider first evaluating the evidence for $M_1$ against $M_2$ based on the results in Tables~\ref{tab:resSBC} and \ref{tab:resSBCinteract} that are based on improper priors. We estimate the likelihood as in Section~\ref{sec:postdev}, relative to the MLE for a model with the contagion parameter, composition, and contagion interaction set to zero. We estimate $\hat{\ell}(\theta_t)$ for a thinned sample of 226 posterior draws, using 20 bridges and 100 samples for each. In fact, using half of these posterior draws and only 5 sampled networks for each give virtually identical results. Figure~\ref{fig:aitkinplot} (left panel) shows that the deviance distributions are stochastically ordered (Aitkin et al., 2017) and that model 2 is the preferred model. Based on the posterior deviances of Figure~\ref{fig:aitkinplot} (left panel), we provide two versions of the DIC measure (Spiegelhalter et al., 2002; Gelman et al., 2004) in Table~\ref{tab:SBCDIC}, both of which suggest that $M_2$ is preferred over $M_1$.

\begin{table}
\caption{\label{tab:SBCDIC}  Deviance Information Criteria (DIC) evaluated for for Models 1 and 2 fitted to the SBC}
\centering
\fbox{%
\begin{tabular}{lrr}
  \hline
 & $\bar{D}+p_D$ &   $\bar{D}+p_V$ \\ 
  \hline
Model 1  & $332.94$ & $333.68$ \\ 
Model 2 & $330.32$ & $329.73$  \\ 
 \hline
\end{tabular}}

\end{table}

Examining the evidence for the two models in Figure~\ref{fig:aitkinplot} (right panel), the interaction model, Model 2, is preferred for $\lambda$ between 1 and 4. As $\lambda$ gets larger, the prior variance increases, penalising model complexity and thus favouring the more parsimonious model (c.p. Bartlett, 1957). The figure is meant to illustrate the dependence on $\lambda$ and the precision of the estimates of the evidence is not sufficient to draw firm conclusions (estimates in the range $\lambda \in (1,3)$ are the average of three estimates with $T=J=2152$, the rest are based on $T=19,000$; a brief illustration of the effect of the sample size is provided in Section F of the Appendix).

\subsubsection{Sensitivity to MAR assumption}
\begin{figure}
\centering
    \makebox{
\includegraphics[scale=.65]{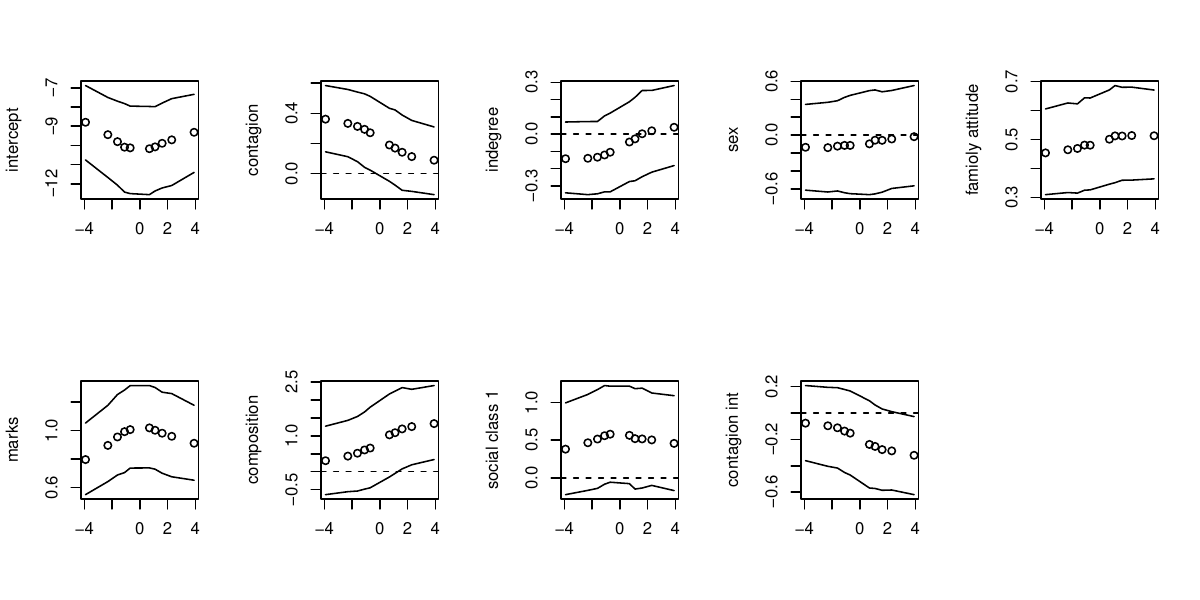}}
\caption{\label{fig:MNARplot}Posterior summaries (95\% CI) for parameters of model 2 (Table~\ref{tab:resSBCinteract}) for SBC against MNAR parameter $\phi_1$}
\end{figure}

To test the sensitivity of the posteriors to violations of the missing at random assumption, we posit the MNAR missing data mechanism assuming the logistic form
\[
\text{logit} \big\{ \Pr(I_i=1|y,x) \big\} = \phi_0+\phi_1y_i+\phi_2 x_{+i} 
\]
independently for all $i\in V$ conditional on $y$. With $\phi_1<0$ the interpretation would be that pupils that do not intend to proceed to higher secondary education are less likely to respond. Assuming that receiving few best-friend nominations is associated with social isolation, a negative $\phi_2$ would mean that socially relatively isolated pupils are more likely to be missing. Fixing $\phi$, only $\phi_1$ will affect inference as the covariate dependent $\phi_2$ and the intercept $\phi_0$ cancel out in simulating $Y_i$ for missing cases. Figure~\ref{fig:MNARplot} plots the change in credibility intervals for some of the parameters of model 2. If missingness is strongly predicted by an intention to proceed to higher secondary education ($\phi_1$ positive), the contagion effect is weakened. If pupils not intending to proceed are more likely to be missing, the contagion effect is strengthened. The strength of the MNAR mechanism also affects the composition parameter and the interaction with composition and contagion. The bias, as represented by $\phi_1$, does however need to be strong to have an effect (at $\phi_1 = 4$, $y_i=1$ almost with probability 1 for missing cases). 
%While missings in $Y$ is not automatically associated with being a non-responder and having missing out-going nominations, pupils with missing on $Y$ nominate fewer people, and if these missings are imputed by $Y=1$, this means increasing the number of isolates that have $Y=1$ when in fact these pupils might have missing values on their out-going ties. 

\subsection{Unemployment in a large network}
When the node-set of a network is not unambiguously defined or the population size is too big to allow for a complete census of the network, we may still want to estimate network-related effects from a sample of the population network.  We consider a dataset analysed previously by Daraganova and Pattison (2013) that consists of 551 individuals recruited via a three-wave snowball sample (Frank, 2005; %Frank \& Snijders, 1994; 
Goodman, 1961) in Australia. Drawing on Besag's (1974) coding scheme, Pattison et al. (2013) demonstrated how the dependence assumptions of an ERGM can be used to define a conditional inference scheme. For ALAAMs this translates to estimating the model as described above with the condition that $y_i$ remains fixed at their observed values for $i\in A \subset V$, where $A$ is a set that \emph{separates} (Pattison et al., 2013) data in $\mathcal{Y}\times\mathcal{X}$. For the three-wave snowball sample this means conditioning on the outcomes of nodes in wave 3 (184 nodes), and conditionally on these nodes, modelling only the outcomes of the seed nodes, and outcomes of nodes in waves 1 and 2 (367 nodes). The outcome variable of interest is employment status with `employed individuals' were those individuals who worked full or part time, and students who worked part time ($Y=0$); and  `unemployed individuals' were those individuals who did not work at the time of the interview ($Y=1$). In addition we use a reduced set of other variables, namely the number of network partners (degree); sex (male: 0; female: 1); and age (ranging from 19 to 67 with a mean of 37).

The results of Table~\ref{tab:gslins551} largely agree with the analysis of Daraganova and Pattison (2013), and there is clear evidence of a positive association between people that are relationally tied (the posterior mean is $0.322$) and a lower risk of being unemployed the more people that you are connected to. Of course, for a sample of a community network we cannot discount the possibility that the network and outcomes are spatially clustered (Butts, 2003; Daraganova et al., 2012%; Koskinen and Lomi, 2013
) or that there are other geographical network effects (Sohn et al., 2019).

%{\textbf{ Galina: I am updating the estimation results of Table~\ref{tab:gslins551}; don't think they will change much but I think sex might be slightly stronger }}
\begin{table}
\caption{\label{tab:gslins551}Posterior summaries for a model for employment status for a sample from Victoria, Australia, estimated conditional on outcomes in waves 3 and greater}
\centering
\fbox{%
\begin{tabular}{lrrrr}
  \hline
 & mean & sd &  2.5 perc & 97.5 perc \\ 
  \hline
intercept	&$	-1.582	$&$	0.246	$&$	-2.119	$&$	-1.13	$\\
contagion	&$	0.324	$&$	0.131	$&$	0.03	$&$	0.548	$\\
degree	&$	-0.106	$&$	0.048	$&$	-0.204	$&$	-0.005	$\\
sex	&$	0.541	$&$	0.264	$&$	0.01	$&$	1.099	$\\
age	&$	0.001	$&$	0.128	$&$	-0.264	$&$	0.242	$\\
   \hline
\end{tabular}}
\end{table}

% Using Pseudo-likelihood rather than real likelihood
%> res.mple$ResTab
%                   mean            sd           ESS       SACF 10       SACF 30
%intercept -1.1847954244  0.2098269209 41.6418492691  0.3059576307 -0.0759870041
%contagion  0.4429098606  0.1186799398 59.2527138541  0.2751801076 -0.0007305562
%          -0.1550588557  0.0466274272 35.0825915757  0.4573912475  0.1912151287
%sex        0.4080730272  0.2372761255 28.0512336860  0.4953046981  0.0270513224
%age       -0.1450692587  0.0914292986 52.5932954473  0.3687968074 -0.0172228822

\section{Summary}
Building on previous work on ALAAMs (Robins et al., 2001; Daraganova, 2009; Daraganova and Robins, 2013) we draw on advances in modelling Markov random fields (Friel, 2013; Caimo and Friel, 2011) to improve on previous Bayesian estimation schemes (Koskinen, 2008) for the social influence model\footnote{Code is available at \url{https://github.com/johankoskinen/ALAAM}}.

We illustrated various aspects of fitting the model using three example datasets. We found that pupils that have friends that have male-dominance attitudes also tend to have male-dominance attitudes themselves. Posterior predictive p-values show that a simple model with direct contagion is sufficient for explaining more complex interactions and in addition show that the ALAAM compares favourably with an alternative network dependence model. For a Swedish dataset we found that pupils that have friends that intend to proceed to higher education are more likely to have the same intention themselves. We also found tentative evidence for this `contagion' effect to be present in schools with pupils of higher social class than in schools with a lower proportion of pupils from a high social class. The estimates for the contagion effect was demonstrated to be robust to violations of the missing at random assumption. Finally, a dataset collected using snowball sampling in Australia showed that people that have unemployed friends are more likely to be unemployed themselves.

A benefit of the Bayesian estimation approach for ALAAMs is that the coherent treatment of uncertainty allows greater flexibility in handling missing data and performing model evaluation relative to the maximum likelihood approach. This likelihood-based framework is also readily extended to hierarchical modelling so that we for example can analyse social influence jointly for multiple datasets (c.p. the continuous case, Agneessens and Koskinen, 2016).
%This multilevel approach would be better suited to investigate network-level effects like the case of differences in contagion for different types of schools.

%Further extensions to the model itself are of course possible and we may consider multiple types of ties, e.g. for independent effects contagion $y_i y_j x_{ij}^{(r)}$ for relations $r=1,\ldots, R$; but we may also consider algebras (Pattison, 1993) on $\prod_{r=1}^{R}\mathcal{X}_r$ in our modelling peer effects - what combinations of different types of social relations are most predictive of an outcome. 

Cross-sectional network data does not allow us to distinguishing social influence and social contagion from \emph{social selection} (Steglich et al., 2010), but when only cross-sectional network data are available it is still necessary to account for peer-dependence through network ties. A Bayesian ALAAM framework allows us to take a number of different types of network dependencies into account. %Investigating social influence, social contagion, and social diffusion has a long tradition of empirical applications in network research (Robins, 2015) and here we have not touched on the potentially rich source of network data that online sources constitute. Indeed, the study of online opinion formation and the spread of fake news through online networks is an increasingly popular field (see e.g., Brady et al., 2017).

%{\bf{ to consider: }}\\
%\emph{ 'simple' and `complex' contagion taken; use `direct' contagion instead;
%Tom wrote something on 2-stage contagion for bi-partite networks;
%think carefully about composition or dummy for SBC;
%try to mention `social statistics' to encourage editor to couch this as a social statistics paper;
%Tom might have written a paper on Social Influecen (check);
%also, check Garry's chapter on social influence;
%for possible reviewers consider `power-users' of RSiena (Schaefer and Scott test); also maybe one of  
% Carter's students. ME: Vitale; Doreian;}

For ERGMs a number of alternatives to the approximate exchange algorithm have been proposed, such as those in Alquirer et al (2016). Given that the dependencies in ALAAMs are considerably weaker than those in ERGMs, approximate algorithms that do not sample from the model hold some promise for ALAAMs.

\section*{Acknowledgements} 
The work by Koskinen was partially supported by NSF-CMMI-2005661 and the  Department of Defense under Grant ARO W911NF-21-1-0335. We are grateful for comments on early drafts by a number of people at IC$^2$S$^2$ and ARS'19, as well as researchers that have patiently tested the inference procedure: Bella Vartanyan, the Science of Networks in Communities (SONIC) research group, and the research group on bushfire data whose work was supported by the grants APP1073041 and the Melbourne climate futures accelerator grant.

\makeatletter
\renewcommand\@biblabel[1]{}
\makeatother

\appendix

\section{The chain graph, moral graph, and non-zero interactions for influence models\label{appendixA}}
To demonstrate the relation between the chain graph and the moral graph, and the probability distribution, we summarise here the proof of the theorem in Robins et al. (2001)\footnote{Henceforth citations refer either to the bibliography of the main article or the one provided here, depending.}, there stated in terms of the conditional distribution of $Y_i=y_i$ given everything else. We use the definition of mutual conditional dependence and the dependence on parent variables in Section 3.1 of the main article.

\vspace{4mm} 
%\begin{theorem}
\noindent\textbf{Theorem}
For parent variables $X$ and child variables $Y$ with chain graph $\mathcal{D}$, given positivity conditions (see below), the conditional distribution of $Y$ given $X=x$ is given by
\begin{equation}
\Pr(Y=y \mid X=x ) =\frac{1}{c}\exp\left\{ \sum_{A 	\subseteq \zeta }\sum_{B 	\subseteq {\mathrm{pa}} (A)  } \gamma_{A \cup B } \prod_{i \in A} y_i \prod_{(j,k) \in B } x_{jk} \right\},
\label{eq:ALAAMmod}
\end{equation}
where $\zeta$ is the set of maximal cliques relating to $Y$, $\gamma_A$ is non-zero if and only if $A$ is a clique in the moral graph obtained from $\mathcal{D}$ by marrying the parents of the same nodes and turning directed ties non-directed. 
%\end{theorem}
\vspace{4mm}

We may prove the claim by a direct application of Hammersley-Clifford theorem for the dependencies implied by $\mathcal{D}$, and then mapping the subsets $A$ that have non-zero interactions to the corresponding cliques in the moral graph. We will see that we do not need to marry parents of the same line-segment as is typically done for chain graphs (Frydenberg, 1990).

For a collection of variables $M$, denote by $M_{i}^{-}$ a collection of variables that is identical to $M$ but where variable $i$ is set to 0. Define $Q(m)=\log \Pr(Y=y \mid X=x)-\log \Pr(Y=0 \mid X=x)$ and we see that for $i \leq n$
\[
Q(m)-Q(m_i^{-}) = \log \left[ \frac{\Pr(Y_i=y_i \mid Y_{-i} = y_{-i} , X = x)}{  \Pr(Y_i=0 \mid Y_{-i} = y_{-i} , X = x) } \right]
\]
and following Besag (1974), we can write $Q(m)$ in terms of its expansion in terms of interactions and functions $\gamma_A$ to obtain
\begin{equation}
Q(m)-Q(m_i^{-}) = m_i \sum_{A \subseteq S \backslash \{i\}} \gamma_{A \cup \{i \}}\prod_{k \in A}m_k,
\label{eq:HCdifff}
\end{equation}
where if $A=\emptyset$ then $\prod_{k\in A}m_k$ is defined as 1.  Based on our definition of mutual conditional dependence and the dependence on parent variables, $Q(m)-Q(m_i^{-})$ can only include $y_i$, $y_j$ where $Y_j$ is a neighbour of $Y_i$, or $x_r$ where $X_r$ is a parent of $Y_i$. 

Pick a set $R \subseteq S$, with $i \in R$ and $i \in V$ and set $m_k=0$ for $k\notin R$, then
\[
Q(m)-Q(m_i^{-}) = y_i \sum_{A \subseteq R \backslash \{i\}} \gamma_{A \cup \{i \}}\prod_{k \in A}m_k.
\]
The set of variables $R$ can be of three forms. Assume $R \subseteq V$ but that $\{ i,j\} \notin \mathcal{D}$, for $j \in R$. Then, $Y_i$ is not mutually conditionally dependent of any $Y_j$ for $j\in R$ and $\gamma_{A \cup \{i\}}=0$ for all $A \subseteq R$. Secondly, assume $R \subseteq T$ but $j \notin \mathrm{pa}(i)$ for $j \in R$, then $\gamma_{A \cup \{i\}}=0$ for all $A \subseteq R$. Thirdly, assume that $R$ contains both parent and child variables and, without loss of generality assume these are $X_h$ and $Y_k$. If $\{ i,k\} \in \mathcal{D}$ but $h \notin \mathrm{pa}(i)$ then $\gamma_{ \{ i,k,h\} }=0$. If  $\{ i,k\} \notin \mathcal{D}$ but $h \in \mathrm{pa}(i)$ then $\gamma_{ \{ i,k,h\} }=0$. Thus  $\gamma_{ \{ i,k,h \} }\neq 0$ if and only if  $\{ i,k\} \in \mathcal{D}$ but $h \in \mathrm{pa}(i)$. Equivalent arguments can be made for $\gamma_{A \cup B}$ for general sets $A \subseteq V$ and $B \subseteq T$. Note that parents of the same line-segment need not be married.

The above requires that $\Pr(Y=y \mid X = x)>0$ for all $y \in \mathcal{Y}$, for each conditioning set $x \in \mathcal{X}$.  While we are not providing any statements about $\Pr(X=x)$ there is no reason to assume that there would be any inherent constraints on the joint sample space $\mathcal{Y} \times \mathcal{X}$ that would lead to a violation of the positivity requirement (Moussouris, 1974).
While the absence of constraints would apply in general, we might be able to construe pathological examples where this is not the case. Assume that we choose to model disease status with ALAAM of a strictly communicable decease, in the sense that with the exception of a seed node, infection status can only be acquired from a network contact. Given knowledge of the seed node and letting $N(i)=\{j \in V: x_{ij}=1 \}$, $\Pr(Y_i = 1 \mid X=x, Y_{N(i)}=0,Y_{-N(i) \cup \{i\}}=y_{-N(i) \cup \{i\}})=0$, violating the positivity assumption. For practical purposes, however, we are never going to have the entire population of susceptible individuals, other than perhaps in fabricated examples, such as the outbreak of a sexually transmissible disease at a polar base.

We may think of a number of ways in which networks are constrained, such as bipartite networks, but constraints on $\mathcal{X}$ do not affect positivity of the conditional distribution. For purposes of deriving non-zero interactions out of the moral graph,  we are free to pick the set $R$ and set $m_k=0$ for $k\notin R$. 

Should we aim to also model $X$, what could we say about the conditional model for $Y$ or even the joint model? While the Markov graph dependence of Frank and Strauss (1986) and the moral graph under [NA] agrees on edges between tie-variables, the more general dependence structure is less straightforward.

\newpage
\section{Partial dependence\label{appendixB}} 
To derive the non-zero interactions from partial dependence structures we follow Pattison and Robins (2002). We note that  Pattison and Robins (2002) define partial dependence in the context of a model for $X$ only but this applies equally to the moral graph of the conditional model $\Pr(Y=y|X=x)$. If the range of the network is full, $\mathcal{X}=\{0.1\}^{ V \choose 2}$, we are free to set $x_i=0$  for any $i \in T$. For partial dependence structures we have the following proposition followed by a corollary:

\vspace{4mm} 
%\begin{prop}
\noindent\textbf{Proposition} 
If $A \subseteq S \backslash B$ and $A$ is not a clique in $\mathcal{Q}_B$ for some subset $B \subset S$, then $\gamma_A=0$ in $\Pr(Y=y|X=x)$.
%\end{prop}
\vspace{4mm} 

%\begin{corollary}
\noindent\textbf{Corollary} 
The $\lambda_A$ interaction is non-zero in the model defined by Eq.~\ref{eq:ALAAMmod} if and only if $A$ is a clique in $\mathcal{Q}$ and in all $\mathcal{Q}_B$ for which $A \cap B = \emptyset$.
%\end{corollary}
\vspace{4mm} 

In Eq.~(\ref{eq:HCdifff}), $\gamma_{A \cup \{ i\}}=0$ unless $A \cup \{ i \}$ is a clique of $\mathcal{Q}$. Now suppose that $C$ is not a clique of $\mathcal{Q}_B$ for some $B$. Assume that  $\{i , j \} \notin \mathcal{Q}_B$ for some $j \in V$  with $i,j \in C \subseteq S \backslash B$. Set $m_k=0$ for all $k \in B$, upon which the function $Q(m)-Q(m_i^{-})$  must be independent of $m_j$. By choosing to set $m_k=0$ for all $k \neq i,j$ we have that $\gamma_{ \{ i,j \} }=0$. If in addition, $h \in C$, we can choose to set $m_k=0$ for all $k \neq i,j,h $, from which it follows that 
\[
Q(m)-Q(m_i^{-}) = \gamma_{ \{ i,j,h\}}m_i m_j m_h =0,
\]
and consequently $ \gamma_{ \{ i,j,h\}}=0$. A similar argument applies to all subsets of $C$.

%{\textbf{Follow Pattison and Robins (2002) where the standard HC theorem is used to show that an interaction is zero unless a clique in the moral graph $\mathcal{T}$; and then that if C is not a clique in the partial graph $\mathcal{Q}_\mathcal{B}$, then interaction is zero}}

\section{Partial dependence structures for direct contagion}

Considering the implied dependence graph, implied by [DA] (as defined in Section 3.1.2 of the main article) for three variables, we obtain the chain graph in Figure~\ref{fig:dep2}(a) with the moral graph in  Figure~\ref{fig:dep2}(b).  The non-zero interactions in the conditional distribution $\Pr(Y=y|X=x)$ are all interactions that are cliques in both the moral graph and all the partial dependence graphs whose node set includes all of the variables of the interaction. For $Y_i$, $Y_j$, and $X_{ij}$, Figure~\ref{fig:partial1} provides a list of the unique (up to isomorphism) partial dependence graphs $\mathcal{Q}_{B}$. In the following, a filled node for $i$ denotes that $y_i=1$, and a non-filled node for $i$ denotes that we do not distinguish between  $y_i=1$ and  $y_i=0$.

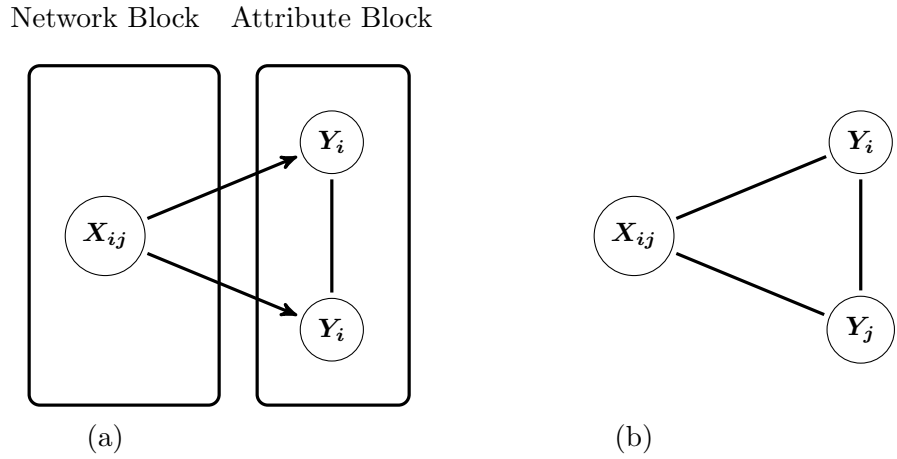
\begin{figure}

\begin{center}
\begin{tikzpicture}
\begin{scope}
\boldmath \tikzstyle{every
node}=[x=3ex,y=2.5ex,shape=circle,minimum size=2ex]
\tikzstyle{every path}=[very thick, -stealth', shorten <=2pt,
shorten >=2pt]

\node (a) at (0,10) {Network Block} ; 
\draw[black,rounded corners] (-1,-1) rectangle (1.5, 3.5) {};
\node[draw,thin] (xij) at (0,3) {{\small $X_{ij}$}} ;

\node (b) at (6,10) {Attribute Block} ; 
\draw[black,rounded corners] (2, -1) rectangle (4, 3.5) {};
\node[draw,thin] (yi) at (6,6) {{\small $Y_{i}$}} ; 
\node[draw,thin] (yj) at (6,0) {{\small $Y_{i}$}} ; 

\draw (xij) -> (yi); \draw (xij) -> (yj);  \draw[-] (yi) -- (yj);

\node[draw,thin] (bxij) at (14,3) {{\small $X_{ij}$}} ;

\node[draw,thin] (byi) at (20,6) {{\small $Y_{i}$}} ; 
\node[draw,thin] (byj) at (20,0) {{\small $Y_{j}$}} ; 

\draw[-] (bxij) -- (byi);\draw[-] (bxij) -- (byj);;\draw[-] (byi) -- (byj);

\node (c) at (0,-3.5) {(a)} ; 
\node (c) at (14,-3.5) {(b)} ;

\end{scope}
\end{tikzpicture}\\
\end{center}
\caption{Dependence graph (a) and Moral graph (b) of network influence model }\label{fig:dep2}
\end{figure}

\begin{figure}
\centering
 \resizebox{380pt}{190pt}{
\begin{tikzpicture}
\begin{scope}
\boldmath \tikzstyle{every
node}=[x=3ex,y=2.5ex,shape=circle,minimum size=2ex]
\tikzstyle{every path}=[very thick, -stealth', shorten <=2pt,
shorten >=2pt]
% titles
\node (moral) at (3,26) {Clique in moral graph} ; 
\node (config) at (15,26) {Configuration} ; 
\node (condit) at (24,26) {$B$} ; 
\node (nodeset) at (30,26) {$S \backslash B$} ; 
\node (part) at (39,26) {Partial dependence graph} ; 
\node (int) at (48,26) {Interaction} ;

% a block - vertical 22
\node (axij) at (0,22) {{\small $X_{ij}$}} ; 
\node  (ayi) at (6,24) {{\small $Y_{i}$}}; 
\node  (ayj) at (6,20) {{\small $Y_{j}$}} ;
\draw[-] (axij) -- (ayi);  \draw[-] (axij) --(ayj);  \draw[-] (ayi) -- (ayj);  

\node[draw,thin,fill=gray] (ai) at (12,22) {{\small $i$}} ; 
\node[draw,thin,fill=gray]  (aj) at (18,22) {{\small $j$}}; 
\draw[-] (ai) -- (aj);  

\node (acondit) at (24,22) {$\emptyset$} ; 
\node (anodeset) at (30,22) {$Y_i,Y_j,X_{ij}$} ; 

\node (aaxij) at (36,22) {{\small $X_{ij}$}} ; 
\node  (aayi) at (42,24) {{\small $Y_{i}$}}; 
\node  (aayj) at (42,20) {{\small $Y_{j}$}} ;
\draw[-] (aaxij) -- (aayi);  \draw[-] (aaxij) --(aayj);  \draw[-] (aayi) -- (aayj);  

\node (aint) at (48,22) {$\neq 0$} ; 

% b block - vertical 16
\node  (byi) at (6,16) {{\small $Y_{i}$}};

\node[draw,thin,fill=gray] (bi) at (12,16) {{\small $i$}} ; 

\node (bcondit) at (24,16) {$Y_j,X_{ij}$} ; 
\node (bnodeset) at (30,16) {$Y_i$} ;

\node  (bbyi) at (42,16) {{\small $Y_{i}$}};

\node (bint) at (48,16) {$\neq 0$} ; 

% c block - vertical 10
\node  (cyi) at (6,12) {{\small $Y_{i}$}}; 
\node  (cyj) at (6,8) {{\small $Y_{j}$}} ;
 \draw[-] (cyi) -- (cyj);  

\node[draw,thin,fill=gray] (ai) at (12,10) {{\small $i$}} ; 
\node[draw,thin,fill=gray]  (aj) at (18,10) {{\small $j$}};

\node (ccondit) at (24,10) {$X_{ij}$} ; 
\node (cnodeset) at (30,10) {$Y_i,Y_j$} ; 

\node  (ccyi) at (42,12) {{\small $Y_{i}$}}; 
\node  (ccyj) at (42,8) {{\small $Y_{j}$}} ;

\node (cint) at (48,10) {$= 0$} ; 

% d  block - vertical 2
\node (dxij) at (0,2) {{\small $X_{ij}$}} ; 
\node  (dyi) at (6,4) {{\small $Y_{i}$}}; 

\draw[-] (dxij) -- (dyi);   

\node[draw,thin,fill=gray] (di) at (12,2) {{\small $i$}} ; 
\node[draw,thin]  (dj) at (18,2) {{\small $j$}}; 
\draw[-] (di) -- (dj);  

\node (dcondit) at (24,2) {$Y_j$} ; 
\node (dnodeset) at (30,2) {$Y_i,X_{ij}$} ; 

\node (ddxij) at (36,2) {{\small $X_{ij}$}} ; 
\node  (ddyi) at (42,4) {{\small $Y_{i}$}}; 

\draw[-] (ddxij) -- (ddyi);
\node (dint) at (48,2) {$\neq 0$} ; 

\end{scope}
\end{tikzpicture}}\\
\caption{Partial dependence graphs and non-zero interactions for a simple network contagion model}\label{fig:partial1}
\end{figure}

%%%%% ===== dependence graph and moral for complex
\section{Partial dependence structures for indirect influence and contagion}\label{sec:appendcomplex}
The [ISI] and [IDA] dependence assumptions (as defined in Section 3.1.3 of the main article, together with the earlier assumptions) yield the directed dependence graph of Figure~\ref{fig:dep2}(a) with moral graph Figure~\ref{fig:dep2}(b).  The sixteen partial dependence graphs on $M=\{Y_i,Y_j,Y_k,X_{ij},X_{jk} \}$ are provided in Figures~\ref{fig:partialcomples1} and ~\ref{fig:partialcomples2}. The resulting statistics with non-zero parameters are depicted in Figure 2 of the main article. Further statistics defined for three outcome variables and three tie-variables include the ones in Figure~\ref{fig:stats3}

\begin{figure}

\begin{center}
\begin{tikzpicture}
\begin{scope}
\boldmath \tikzstyle{every
node}=[x=3ex,y=2.5ex,shape=circle,minimum size=2ex]
\tikzstyle{every path}=[very thick, -stealth', shorten <=2pt,
shorten >=2pt]

\node (a) at (0,10) {Network Block} ; 
\draw[black,rounded corners] (-1, -1) rectangle (1.5, 3.5) {};
\node[draw,thin] (xij) at (0,5) {{\small $X_{ij}$}} ; 
\node[draw,thin]  (xjk) at (0,1) {{\small $X_{jk}$}};

\node (b) at (6,10) {Attribute Block} ; 
\draw[black,rounded corners] (2, -1) rectangle (4, 3.5) {};
\node[draw,thin] (yi) at (6,6) {{\small $Y_{i}$}} ; 
\node[draw,thin] (yj) at (7.5,3) {{\small $Y_{j}$}} ; 
\node[draw,thin] (yk) at (6,0) {{\small $Y_{k}$}} ; 

\draw (xij) -> (yi); \draw (xij) -> (yj); \draw (xij) -> (yk); 
\draw (xjk) -> (yi); \draw (xjk) -> (yj); \draw (xjk) -> (yk); 

\draw[-] (yi) -- (yj);  \draw[-] (yi) -- (yk); \draw[-] (yj) -- (yk);

\node[draw,thin] (bxij) at (14,5) {{\small $X_{ij}$}} ; 
\node[draw,thin]  (bxjk) at (14,1) {{\small $X_{jk}$}};

\node[draw,thin] (byi) at (20,6) {{\small $Y_{i}$}} ; 
\node[draw,thin] (byj) at (21.5,3) {{\small $Y_{j}$}} ; 
\node[draw,thin] (byk) at (20,0) {{\small $Y_{k}$}} ; 

\draw[-] (bxij) -- (byi); \draw[-] (bxij) -- (byj);\draw[-] (bxij) -- (byk);
\draw[-] (bxjk) -- (byi);\draw[-] (bxjk) -- (byj);\draw[-] (bxjk) -- (byk);
\draw[-] (bxij) -- (bxjk); 
\draw[-] (byi) -- (byj);  \draw[-] (byi) -- (byk); \draw[-] (byj) -- (byk);   

\node (c) at (0,-3.5) {(a)} ; 
\node (c) at (14,-3.5) {(b)} ;

\end{scope}
\end{tikzpicture}\\
\end{center}
\caption{Dependence graph (a) and Moral graph (b) of complex contagion model (Robins et al., 2001)}\label{fig:dep2}
\end{figure}

\begin{figure}
\centering
 \resizebox{380pt}{300pt}{
\begin{tikzpicture}
\begin{scope}
\boldmath \tikzstyle{every
node}=[x=3ex,y=2.5ex,shape=circle,minimum size=2ex]
\tikzstyle{every path}=[very thick, -stealth', shorten <=2pt,
shorten >=2pt]
% titles
\node (moral) at (3,46) {Clique in moral graph} ; 
\node (config) at (15,46) {Configuration} ; 
\node (condit) at (24,46) {$B$} ; 
\node (nodeset) at (30,46) {$S \backslash B$} ; 
\node (part) at (39,46) {Partial dependence graph} ; 
\node (int) at (48,46) {Interaction} ;

%%  a block - vertical 42 - %%%%
% moral graph
\node (axij) at (0,43.5) {{\small $X_{ij}$}} ; 
\node (axjk) at (0,40) {{\small $X_{jk}$}} ; 
\node  (ayi) at (4,44) {{\small $Y_{i}$}}; 
\node  (ayj) at (6,42) {{\small $Y_{j}$}} ;
\node  (ayk) at (4,40) {{\small $Y_{k}$}} ;
\draw[ -stealth', shorten <=2pt, shorten >=2pt, -] (axij) to [bend right] (axjk); \draw[-] (axij) -- (ayi);  \draw[-] (axij) --(ayj);  \draw[-] (axij) --(ayk); %x_ij ->
\draw[-] (axjk) -- (ayi);  \draw[-] (axjk) --(ayj);  \draw[-] (axjk) --(ayk); % X_jk ->
\draw[-] (ayi) -- (ayj);  \draw[-] (ayi) -- (ayk);  % y_i ->
 \draw[-] (ayj) -- (ayk);  
% configuaration

\node[draw,thin,fill=gray] (ai) at (12,42) {{\small $i$}} ; 
\node[draw,thin,fill=gray]  (aj) at (15,42) {{\small $j$}}; 
\node[draw,thin,fill=gray]  (ak) at (18,42) {{\small $k$}}; 
\draw[-] (ai) -- (aj);  \draw[-] (aj) -- (ak);  
% conditioning
\node (acondit) at (24,42) {$\emptyset$} ; 
% nodeset
\node (anodeset) at (30,42) {$Y_i,Y_j,Y_k,X_{ij},X_{jk}$} ; 
% partial dependence graph
\node (aaxij) at (36,43.5) {{\small $X_{ij}$}} ; 
\node (aaxjk) at (36,40) {{\small $X_{jk}$}} ; 
\node  (aayi) at (40,44) {{\small $Y_{i}$}}; 
\node  (aayj) at (42,42) {{\small $Y_{j}$}} ;
\node  (aayk) at (40,40) {{\small $Y_{k}$}} ;
\draw[-] (aaxij) to [bend right] (aaxjk); 
\draw[-] (aaxij) -- (aayi);  \draw[-] (aaxij) --(aayj);  \draw[-] (aaxij) --(aayk);  %x_ij ->
\draw[-] (aaxjk) -- (aayi);  \draw[-] (aaxjk) --(aayj);  \draw[-] (aaxjk) --(aayk);  % X_jk ->
\draw[-] (aayi) -- (aayj);  \draw[-] (aayi) -- (aayk);   % y_i ->
 \draw[-] (aayj) -- (aayk);  

%parameter

\node (aint) at (48,42) {$\neq 0$} ; 

% b block - vertical 36
\node  (byi) at (4,36) {{\small $Y_{i}$}}; 

% configuaration
\node[draw,thin,fill=gray] (bi) at (12,36) {{\small $i$}} ; 
% conditioning
\node (bcondit) at (24,36) {$Y_j,Y_k,X_{ij},X_{jk}$} ; 
% nodeset
\node (bnodeset) at (30,36) {$Y_i$} ; 

% partial dependence graph
\node  (bbyi) at (40,36) {{\small $Y_{i}$}}; 

%parameter

\node (bint) at (48,36) {$\neq 0$} ; 

%%% c block - vertical 30 - %%%
% moral graph
\node (cxjk) at (0,29) {{\small $X_{jk}$}} ; 
\node  (cyi) at (4,32) {{\small $Y_{i}$}}; 
\draw[-] (cxjk) -- (cyi);   

% configuaration

\node[draw,thin,fill=gray] (ci) at (12,30) {{\small $i$}} ; 
\node[draw,thin]  (cj) at (15,30) {{\small $j$}}; 
\node[draw,thin]  (ck) at (18,30) {{\small $k$}}; 
 \draw[-] (cj) -- (ck);  

% conditioning
\node (ccondit) at (24,30) {$Y_j,Y_k,X_{ij}$} ; 

% nodeset
\node (anodeset) at (30,30) {$Y_i,X_{jk}$} ; 
% partial dependence graph
\node (ccxjk) at (36,29) {{\small $X_{jk}$}} ; 
\node  (ccyi) at (40,32) {{\small $Y_{i}$}}; 

%parameter

\node (cint) at (48,30) {$= 0$} ; 

%%  d block - vertical 24 - %%%%
% moral graph
\node (dxij) at (0,23) {{\small $X_{ij}$}} ; 
\node  (dyi) at (4,24) {{\small $Y_{i}$}}; 
 \draw[-] (dxij) -- (dyi);
% configuaration

\node[draw,thin,fill=gray] (di) at (12,24) {{\small $i$}} ; 
\node[draw,thin]  (dj) at (15,24) {{\small $j$}}; 
\draw[-] (di) -- (dj); 
% conditioning
\node (dcondit) at (24,24) {$Y_j,Y_k,X_{jk}$} ; 
% nodeset
\node (dnodeset) at (30,24) {$Y_i,X_{ij}$} ; 
% partial dependence graph
\node (ddxij) at (36,23) {{\small $X_{ij}$}} ; 
\node  (ddyi) at (40,24) {{\small $Y_{i}$}}; 
\draw[-] (ddxij) -- (ddyi); 

%parameter

\node (dint) at (48,24) {$\neq 0$} ; 

%%  e block - vertical 18 - %%%%
% moral graph

\node  (eyi) at (4,20) {{\small $Y_{i}$}}; 
\node  (eyk) at (4,16) {{\small $Y_{k}$}} ;
 \draw[-] (eyi) -- (eyk);  % y_i ->
% configuaration

\node[draw,thin,fill=gray] (ei) at (12,18) {{\small $i$}} ; 
\node[draw,thin,fill=gray]  (ek) at (18,18) {{\small $k$}}; 

% conditioning
\node (econdit) at (24,18) {$Y_j,X_{ij},X_{jk}$} ; 
% nodeset
\node (enodeset) at (30,18) {$Y_i,Y_k$} ; 
% partial dependence graph
\node  (eeyi) at (40,20) {{\small $Y_{i}$}}; 
\node (eeyk) at (40,16) {{\small $Y_{k}$}} ;

%parameter

\node (eint) at (48,18) {$= 0$} ; 

%%%%%%
%%  f block - vertical 10 - %%%%
% moral graph

\node  (fyi) at (4,12) {{\small $Y_{i}$}}; 
\node  (fyj) at (6,10) {{\small $Y_{j}$}} ;

\draw[-] (fyi) -- (fyj);
% configuaration

\node[draw,thin,fill=gray] (fi) at (12,10) {{\small $i$}} ; 
\node[draw,thin,fill=gray]  (fj) at (15,10) {{\small $j$}}; 

% conditioning
\node (fcondit) at (24,10) {$Y_k,X_{ij},X_{jk}$} ; 
% nodeset
\node (fnodeset) at (30,10) {$Y_i,Y_j$} ; 
% partial dependence graph
\node  (ffyi) at (40,12) {{\small $Y_{i}$}}; 
\node  (ffyj) at (42,10) {{\small $Y_{j}$}} ;

%parameter

\node (fint) at (48,10) {$= 0$} ; 
%%%%%%
%%  g block - vertical 4 - %%%%
% moral graph
\node (gxjk) at (0,3) {{\small $X_{jk}$}} ; 
\node  (gyi) at (4,6) {{\small $Y_{i}$}}; 
\node  (gyk) at (4,2) {{\small $Y_{k}$}} ;
\draw[-] (gxjk) -- (gyi);   \draw[-] (gxjk) --(gyk); % X_jk ->
 \draw[-] (gyi) -- (gyk);  % y_i ->
% configuaration

\node[draw,thin,fill=gray] (gi) at (12,4) {{\small $i$}} ; 
\node[draw,thin]  (gj) at (15,4) {{\small $j$}}; 
\node[draw,thin,fill=gray]  (gk) at (18,4) {{\small $k$}}; 
 \draw[-] (gj) -- (gk);  
% conditioning
\node (gcondit) at (24,4) {$Y_j,X_{ij}$} ; 
% nodeset
\node (gnodeset) at (30,4) {$Y_i,Y_k,X_{jk}$} ; 
% partial dependence graph
\node (ggxjk) at (36,3) {{\small $X_{jk}$}} ; 
\node  (ggyi) at (40,6) {{\small $Y_{i}$}}; 
\node  (ggyk) at (40,2) {{\small $Y_{k}$}} ;
 \draw[-] (ggxjk) --(ggyk);  % X_jk ->

%parameter

\node (gint) at (48,4) {$=0$} ; 

%%%%%%

%%  h block - vertical -4 - %%%%
% moral graph
\node (hxjk) at (0,-5) {{\small $X_{jk}$}} ; 
\node  (hyi) at (4,-2) {{\small $Y_{i}$}}; 
\node  (hyj) at (6,-4) {{\small $Y_{j}$}} ;
\draw[-] (hxjk) -- (hyi);  \draw[-] (hxjk) --(hyj); 
\draw[-] (hyi) -- (hyj); 
% configuaration

\node[draw,thin,fill=gray] (hi) at (12,-4) {{\small $i$}} ; 
\node[draw,thin,fill=gray]  (hj) at (15,-4) {{\small $j$}}; 
\node[draw,thin]  (hk) at (18,-4) {{\small $k$}}; 
 \draw[-] (hj) -- (hk);  
% conditioning
\node (hcondit) at (24,-4) {$Y_k,X_{ij}$} ; 
% nodeset
\node (hnodeset) at (30,-4) {$Y_i,Y_j,X_{jk}$} ; 
% partial dependence graph
\node (hhxjk) at (36,-5) {{\small $X_{jk}$}} ; 
\node  (hhyi) at (40,-2) {{\small $Y_{i}$}}; 
\node  (hhyj) at (42,-4) {{\small $Y_{j}$}} ;
 \draw[-] (hhxjk) --(hhyj);

%parameter

\node (aint) at (48,-4) {$=0$} ;

%%%%%%
\end{scope}
\end{tikzpicture}}\\
\caption{Partial dependence graphs and non-zero interactions for dependence assumptions for complex contagion model (part 1)}\label{fig:partialcomples1}
\end{figure}

%%%%%%%%%%%%%%%%%%%%%%%%%%%%
%% 
%% Second PICTURE
%%
%%%%%%%%%%%%%%%%%%%%%%%%%%%%

\begin{figure}
\centering
 \resizebox{380pt}{380pt}{
\begin{tikzpicture}
\begin{scope}
\boldmath \tikzstyle{every
node}=[x=3ex,y=2.5ex,shape=circle,minimum size=2ex]
\tikzstyle{every path}=[very thick, -stealth', shorten <=2pt,
shorten >=2pt]
% titles
\node (moral) at (3,46) {Clique in moral graph} ; 
\node (config) at (15,46) {Configuration} ; 
\node (condit) at (24,46) {$B$} ; 
\node (nodeset) at (30,46) {$S \backslash B$} ; 
\node (part) at (39,46) {Partial dependence graph} ; 
\node (int) at (48,46) {Interaction} ;
\node (axij) at (0,43.5) {{\small $X_{ij}$}} ; 
\node  (ayi) at (4,44) {{\small $Y_{i}$}}; 
\node  (ayk) at (4,40) {{\small $Y_{k}$}} ;
 \draw[-] (axij) -- (ayi);  \draw[-] (axij) --(ayk); %x_ij ->
 \draw[-] (ayi) -- (ayk);  % y_i ->

% configuaration

\node[draw,thin,fill=gray] (ai) at (12,42) {{\small $i$}} ; 
\node[draw,thin]  (aj) at (15,42) {{\small $j$}}; 
\node[draw,thin,fill=gray]  (ak) at (18,42) {{\small $k$}}; 
\draw[-] (ai) -- (aj); 
% conditioning
\node (acondit) at (24,42) {$Y_j,X_{jk}$} ; 
% nodeset
\node (anodeset) at (30,42) {$Y_i,Y_k,X_{ij}$} ; 
% partial dependence graph
\node (aaxij) at (36,43.5) {{\small $X_{ij}$}} ; 
\node  (aayi) at (40,44) {{\small $Y_{i}$}}; 
\node  (aayk) at (40,40) {{\small $Y_{k}$}} ;
\draw[-] (aaxij) -- (aayi);

%parameter

\node (aint) at (48,42) {$= 0$} ; 

%%%%%%%%%%%%
% b block - vertical 34
\node (bxij) at (0,35) {{\small $X_{ij}$}} ; 
\node  (byi) at (4,36) {{\small $Y_{i}$}}; 
\node  (byj) at (6,34) {{\small $Y_{j}$}} ;
\draw[-] (bxij) -- (byi);  \draw[-] (bxij) --(byj); 
\draw[-] (byi) -- (byj); 

%% configuaration
%
\node[draw,thin,fill=gray] (bi) at (12,34) {{\small $i$}} ; 
\node[draw,thin,fill=gray]  (bj) at (15,34) {{\small $j$}}; 

\draw[-] (bi) -- (bj); 
%% conditioning
\node (bcondit) at (24,34) {$Y_k.X_{jk}$} ; 
%% nodeset
\node (bnodeset) at (30,34) {$Y_i,Y_j,X_{ij}$} ; 
%% partial dependence graph
\node (bbxij) at (36,35) {{\small $X_{ij}$}} ; 
\node  (byi) at (40,36) {{\small $Y_{i}$}}; 
\node  (bbyj) at (42,34) {{\small $Y_{j}$}} ;
\draw[-] (bbxij) -- (bbyi);  \draw[-] (bbxij) --(bbyj);
\draw[-] (bbyi) -- (bbyj); 

%parameter

\node (bint) at (48,34) {$\neq 0$} ;

%%  c block - vertical 28 - %%%%
% moral graph
\node (cxij) at (0,29.5) {{\small $X_{ij}$}} ; 
\node (cxjk) at (0,26) {{\small $X_{jk}$}} ; 
\node  (cyi) at (4,30) {{\small $Y_{i}$}}; 
\draw[-] (cxij) to [bend right] (cxjk); \draw[-] (cxij) -- (cyi);
\draw[-] (cxjk) -- (cyi); 
%% configuaration
%
\node[draw,thin,fill=gray] (ci) at (12,28) {{\small $i$}} ; 
\node[draw,thin]  (cj) at (15,28) {{\small $j$}}; 
\node[draw,thin]  (ck) at (18,28) {{\small $k$}}; 
\draw[-] (ci) -- (cj);  \draw[-] (cj) -- (ck);  
%% conditioning
\node (ccondit) at (24,28) {$Y_k,Y_j$} ; 
%% nodeset
\node (cnodeset) at (30,28) {$Y_i,X_{ij},X_{jk}$} ; 
%% partial dependence graph
\node (ccxij) at (36,29.5) {{\small $X_{ij}$}} ; 
\node (ccxjk) at (36,26) {{\small $X_{jk}$}} ; 
\node  (ccyi) at (40,30) {{\small $Y_{i}$}}; 

\draw[-] (ccxij) to [bend right] (ccxjk); 
\draw[-] (ccxij) -- (ccyi);
\draw[-] (ccxjk) -- (ccyi); 

%parameter

\node (cint) at (48,28) {$\neq 0$} ; 

%%%%%%%%%%%%%%%%%
%%  d  block - vertical 20 - %%%%
% moral graph
\node  (dyi) at (4,22) {{\small $Y_{i}$}}; 
\node  (dyj) at (6,20) {{\small $Y_{j}$}} ;
\node  (dyk) at (4,18) {{\small $Y_{k}$}} ;

\draw[-] (dyi) -- (dyj);  \draw[-] (dyi) -- (dyk);  % y_i ->
 \draw[-] (dyj) -- (dyk);  
%% configuaration
%
\node[draw,thin,fill=gray] (di) at (12,20) {{\small $i$}} ; 
\node[draw,thin,fill=gray]  (dj) at (15,20) {{\small $j$}}; 
\node[draw,thin,fill=gray]  (dk) at (18,20) {{\small $k$}}; 

%% conditioning
\node (dcondit) at (24,20) {$X_{ik},X_{jk}$} ; 
%% nodeset
\node (dnodeset) at (30,20) {$Y_i,Y_j,Y_k$} ; 
%% partial dependence graph

\node  (ddyi) at (40,22) {{\small $Y_{i}$}}; 
\node  (ddyj) at (42,20) {{\small $Y_{j}$}} ;
\node  (ddyk) at (40,18) {{\small $Y_{k}$}} ;

%parameter

\node (dint) at (48,22) {$= 0$} ; 

%%%%%%%%%%%%%%%%%
%%  e block - vertical 12 - %%%%
% moral graph
\node (exij) at (0,13.5) {{\small $X_{ij}$}} ; 
\node (exjk) at (0,10) {{\small $X_{jk}$}} ; 
\node  (eyi) at (4,14) {{\small $Y_{i}$}}; 
\node  (eyk) at (4,10) {{\small $Y_{k}$}} ;
\draw[-] (exij) to [bend right] (exjk); \draw[-] (exij) -- (eyi); \draw[-] (exij) --(eyk); %x_ij ->
\draw[-] (exjk) -- (eyi);   \draw[-] (exjk) --(eyk); % X_jk ->
 \draw[-] (eyi) -- (eyk);  % y_i ->

%% configuaration
%
\node[draw,thin,fill=gray] (ei) at (12,12) {{\small $i$}} ; 
\node[draw,thin]  (ej) at (15,12) {{\small $j$}}; 
\node[draw,thin,fill=gray]  (ek) at (18,12) {{\small $k$}}; 
\draw[-] (ei) -- (ej);  \draw[-] (ej) -- (ek);  
%% conditioning
\node (econdit) at (24,12) {$Y_j$} ; 
%% nodeset
\node (enodeset) at (30,12) {$Y_i,Y_k,X_{ij},X_{jk}$} ; 
%% partial dependence graph
\node (eexij) at (36,13.5) {{\small $X_{ij}$}} ; 
\node (eexjk) at (36,10) {{\small $X_{jk}$}} ; 
\node  (eeyi) at (40,14) {{\small $Y_{i}$}}; 
\node  (eeyk) at (40,10) {{\small $Y_{k}$}} ;
\draw[-] (eexij) to [bend right] (eexjk); 
\draw[-] (eexij) -- (eeyi);    \draw[-] (eexij) --(eeyk);  %x_ij ->
\draw[-] (eexjk) -- (eeyi);  \draw[-] (eexjk) --(eeyk);  % X_jk ->
  \draw[-] (eeyi) -- (eeyk);   % y_i ->

%parameter

\node (eint) at (48,12) {$\neq 0$} ; 

%%%%%%%%%%%%%%%%%
%%  f block - vertical 4 - %%%%
% moral graph
\node (fxij) at (0,5.5) {{\small $X_{ij}$}} ; 
\node (fxjk) at (0,2) {{\small $X_{jk}$}} ; 
\node  (fyi) at (4,6) {{\small $Y_{i}$}}; 
\node  (fyj) at (6,4) {{\small $Y_{j}$}} ;

\draw[-] (fxij) to [bend right] (fxjk); \draw[-] (fxij) -- (fyi);  \draw[-] (fxij) --(fyj); 
\draw[-] (fxjk) -- (fyi);  \draw[-] (fxjk) --(fyj); 
\draw[-] (fyi) -- (fyj); 
%% configuaration
%
\node[draw,thin,fill=gray] (fi) at (12,4) {{\small $i$}} ; 
\node[draw,thin,fill=gray]  (fj) at (15,4) {{\small $j$}}; 
\node[draw,thin]  (fk) at (18,4) {{\small $k$}}; 
\draw[-] (fi) -- (fj);  \draw[-] (fj) -- (fk);  
%% conditioning
\node (fcondit) at (24,4) {$Y_k$} ; 
%% nodeset
\node (fnodeset) at (30,4) {$Y_i,Y_j,X_{ij},X_{jk}$} ; 
%% partial dependence graph
\node (ffxij) at (36,5.5) {{\small $X_{ij}$}} ; 
\node (ffxjk) at (36,2) {{\small $X_{jk}$}} ; 
\node  (ffyi) at (40,6) {{\small $Y_{i}$}}; 
\node  (ffyj) at (42,4) {{\small $Y_{j}$}} ;

\draw[-] (ffxij) to [bend right] (ffxjk); 
\draw[-] (ffxij) -- (ffyi);  \draw[-] (ffxij) --(ffyj);
\draw[-] (ffxjk) -- (ffyi);  \draw[-] (ffxjk) --(ffyj); 
\draw[-] (ffyi) -- (ffyj);

%parameter

\node (fint) at (48,4) {$\neq 0$} ; 

%%%%%%%%%%%%%%%%%
%%  g block - vertical -2 - %%%%
% moral graph
\node (gxij) at (0,-1) {{\small $X_{ij}$}} ; 
\node  (gyi) at (4,0) {{\small $Y_{i}$}}; 
\node  (gyj) at (6,-2) {{\small $Y_{j}$}} ;
\node  (gyk) at (4,-4) {{\small $Y_{k}$}} ;
 \draw[-] (gxij) -- (gyi);  \draw[-] (gxij) --(gyj);  \draw[-] (gxij) --(gyk); %x_ij ->
\draw[-] (gyi) -- (gyj);  \draw[-] (gyi) -- (gyk);  % y_i ->
 \draw[-] (gyj) -- (gyk);  
%% configuaration
%
\node[draw,thin,fill=gray] (gi) at (12,-2) {{\small $i$}} ; 
\node[draw,thin,fill=gray]  (gj) at (15,-2) {{\small $j$}}; 
\node[draw,thin,fill=gray]  (gk) at (18,-2) {{\small $k$}}; 
\draw[-] (gi) -- (gj);
%% conditioning
\node (gcondit) at (24,-2) {$X_{jk}$} ; 
%% nodeset
\node (gnodeset) at (30,-2) {$Y_i,Y_j,Y_k,X_{ij}$} ; 
%% partial dependence graph
\node (ggxij) at (36,-1) {{\small $X_{ij}$}} ; 
\node  (ggyi) at (40,0) {{\small $Y_{i}$}}; 
\node  (ggyj) at (42,-2) {{\small $Y_{j}$}} ;
\node  (ggyk) at (40,-4) {{\small $Y_{k}$}} ;

\draw[-] (ggxij) -- (ggyi);  \draw[-] (ggxij) --(ggyj); 
\draw[-] (ggyi) -- (ggyj); 

%parameter

\node (gint) at (48,-2) {$= 0$} ; 

%%%%%%%%%%%%%%%%%
%%  h block - vertical -10 - %%%%
% moral graph

\node (hxjk) at (0,-11.5) {{\small $X_{jk}$}} ; 
\node  (hyi) at (4,-8) {{\small $Y_{i}$}}; 
\node  (hyj) at (6,-10) {{\small $Y_{j}$}} ;
\node  (hyk) at (4,-12) {{\small $Y_{k}$}} ;
\draw[-] (hxjk) -- (hyi);  \draw[-] (hxjk) --(hyj);  \draw[-] (hxjk) --(hyk); % X_jk ->
\draw[-] (hyi) -- (hyj);  \draw[-] (hyi) -- (hyk);  % y_i ->
 \draw[-] (hyj) -- (hyk);  
%% configuaration
%
\node[draw,thin,fill=gray] (hi) at (12,-10) {{\small $i$}} ; 
\node[draw,thin,fill=gray]  (hj) at (15,-10) {{\small $j$}}; 
\node[draw,thin,fill=gray]  (hk) at (18,-10) {{\small $k$}}; 
 \draw[-] (hj) -- (hk);  
%% conditioning
\node (hcondit) at (24,-10) {$X_{ij}$} ; 
%% nodeset
\node (hnodeset) at (30,-10) {$Y_i,Y_j,Y_k,X_{jk}$} ; 
%% partial dependence graph

\node (hhxjk) at (36,-11.5) {{\small $X_{jk}$}} ; 
\node  (hhyi) at (40,-8) {{\small $Y_{i}$}}; 
\node  (hhyj) at (42,-10) {{\small $Y_{j}$}} ;
\node  (hhyk) at (40,-12) {{\small $Y_{k}$}} ;

 \draw[-] (hhxjk) --(hhyk);  % X_jk ->
  \draw[-] (hhyj) -- (hhyk);   % y_i ->

%parameter

\node (hint) at (48,-10) {$= 0$} ; 

%%%%%%%%%%%%%%%%%
%%%%%%
\end{scope}
\end{tikzpicture}}\\
\caption{Partial dependence graphs and non-zero interactions for dependence assumptions for complex contagion model (part 2)}\label{fig:partialcomples2}
\end{figure}

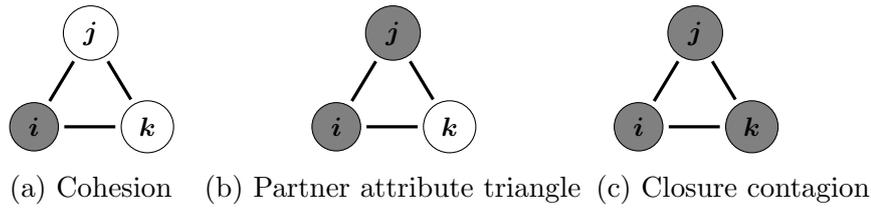
\begin{figure}

\begin{center}
\begin{tikzpicture}
\begin{scope}
\boldmath \tikzstyle{every
node}=[x=3ex,y=2.5ex,shape=circle,minimum size=2ex]
\tikzstyle{every path}=[very thick, -stealth', shorten <=2pt,
shorten >=2pt]

%%% === a === %%%%
\node (a) at (1.5,0) {(a)  Cohesion};
\node[draw,thin,fill=gray] (ai) at (0,2) {{\small $i$}} ; 
\node[draw,thin] (aj) at (1.5,5) {{\small $j$}} ; 
\node[draw,thin] (ak) at (3,2) {{\small $k$}} ; 
 \draw[-] (ai) -- (aj); 
 \draw[-] (ai) -- (ak);
 \draw[-] (ak) -- (aj);  
%%% === b === %%%%
\node (b) at (9.5,0) {(b)  Partner attribute triangle};
\node[draw,thin,fill=gray] (bi) at (8,2) {{\small $i$}} ; 
\node[draw,thin,fill=gray] (bj) at (9.5,5) {{\small $j$}} ; 
\node[draw,thin] (bk) at (11,2) {{\small $k$}} ; 
 \draw[-] (bi) -- (bj); 
 \draw[-] (bi) -- (bk);
 \draw[-] (bk) -- (bj);  

%%% === c === %%%%
\node (c) at (18.5,0) {(c) Closure contagion};
\node[draw,thin,fill=gray] (ci) at (16,2) {{\small $i$}} ; 
\node[draw,thin,fill=gray] (cj) at (17.5,5) {{\small $j$}} ; 
\node[draw,thin,fill=gray] (ck) at (19,2) {{\small $k$}} ; 
 \draw[-] (ci) -- (cj); 
 \draw[-] (ci) -- (ck);
 \draw[-] (ck) -- (cj);  
\end{scope}
\end{tikzpicture}\\
\end{center}
\caption{Statistics associated with the indirect structural dependence assumption and the indirect dependent attribute assumption }\label{fig:stats3}
\end{figure}

\newpage
\section{Directed graphs}
The method for deriving non-zero interactions above is agnostic as to whether $G$ is directed or un-directed. However, care has to be taken to make sure that the dependence structure is coherent. Consider for example the dependence assumption for direct contagion. While a node $i$ may have a tie to node $j$, $x_{ij}=1$ as well as a tie from $j$, $x_{ji}=1$, if $Y_i$ may be conditionally dependent on $Y_j$ if  $x_{ij}=1$, then  $Y_i$  must be allowed to be conditionally dependent on $Y_i$ whenever $x_{ij}=1$ as well as when $x_{ji}=1$. The cliques in the partial dependence graphs $\mathcal{Q}_{B}$ and $\mathcal{Q}_{B^{\prime}}$ induced by $B=S\backslash \{ Y_i,Y_j,X_{ij} \}$ and  $B^{\prime}=S\backslash \{ Y_i,Y_j,X_{ji} \}$, respectively, are distinct but the corresponding statistics are isomorphic. The dependence assumption for direct contagion for directed graphs must thus be that $Y_i$ and $Y_j$  are conditionally dependent if and only if $x_{ij}=1$ or $x_{ji}=1$. This dependence assumption also yields a statistic for reciprocal contagion with statistics of the kind $Y_iY_j X_{ij}X_{ji}$.

\newpage
\section{Variance of marginal likelihood}\label{sec:appendevid}
For the example in Section 5.2 of the main article, Table~\ref{tab:evidencesd} compares the estimates and standard deviation for the marginal likelihood for Model 1 for different choices of $T$, where $T=J$. The number of graphs drawn from the model is $G=M=100$. The likelihood has been evaluated using the path sampler with $20$ bridges and the same estimated likelihood is used for each value of $\lambda$.

\begin{table}
\caption{\label{tab:evidencesd}Calculations of the marginal likelihood for Model 1 in Table 4 in main article}
\centering
\fbox{%
\begin{tabular}{rrrr}
  \hline
 $\lambda$ & $T$ & mean & sd \\ 
  \hline
1 & 2400 & -210.00 & 0.26 \\ 
 & 4800 & -209.98 & 0.26 \\ 
 & 6300 & -209.94 & 0.11 \\ 
2 & 2400 & -206.59 & 0.14 \\ 
 & 4800 & -206.58 & 0.10 \\ 
 & 6300 & -206.57 & 0.12 \\ 
4& 2400 & -191.91 & 0.22 \\ 
 & 4800 & -191.91 & 0.15 \\ 
 & 6300 & -191.92 & 0.12 \\ 
16& 2400 & -178.28 & 0.43 \\ 
 & 4800 & -178.19 & 0.30 \\ 
 & 6300 & -178.20 & 0.20 \\ 
32 & 2400 & -174.78 & 0.44 \\ 
 & 4800 & -174.83 & 0.35 \\ 
 & 6300 & -174.90 & 0.24 \\ 
    \hline
 \end{tabular}}
 \end{table}

\makeatletter
\renewcommand\@biblabel[1]{}
\makeatother

\end{document}